\documentclass[12pt,oneside,onecolumn,floatfix]{article}
\usepackage{epsfig}
\usepackage{a4wide}
\usepackage[top=30pt,bottom=30pt,left=48pt,right=46pt]{geometry}
\usepackage{slashed, enumitem}
\usepackage[utf8]{inputenc}
\usepackage{jcappub}
\usepackage{float,color}
\usepackage{subfigure}
\usepackage{multirow}
\usepackage{slashed}
\usepackage{placeins}
\usepackage{wasysym} 
\usepackage{mathrsfs} 
\usepackage{caption}
\usepackage{amsmath}
\usepackage{amssymb}
\usepackage{graphicx}
\usepackage{slashed}
\usepackage{multirow}
\usepackage{placeins}
\usepackage[dvipsnames]{xcolor}
\usepackage{epstopdf}
\usepackage{soul}
\usepackage{tikz}
\usetikzlibrary{matrix, positioning}
\usepackage[capitalise, english]{cleveref}
\usepackage{siunitx}
\usepackage{xspace}
\usetikzlibrary{trees}
\usetikzlibrary{decorations.pathmorphing}
\usetikzlibrary{decorations.markings}
\usepackage[colorlinks=true,citecolor=blue,linkcolor=blue]{hyperref}
\usepackage{mhchem}
\usepackage{upgreek}
\usepackage{mathtools}
\usepackage{setspace, slashed, gensymb, hhline, booktabs}
\usepackage{multirow}
\usepackage{placeins}
\usepackage[dvipsnames]{xcolor}
\usepackage{epstopdf}
\usepackage{soul}
\usepackage{tikz}
\usetikzlibrary{positioning}
\usetikzlibrary{shapes.geometric, arrows}
\usepackage[capitalise, english]{cleveref}
\usepackage{siunitx}
\usepackage{xspace}
\usepackage{booktabs}
\usetikzlibrary{trees}
\usetikzlibrary{decorations.pathmorphing}
\usetikzlibrary{decorations.markings}

\usepackage{pifont}% http://ctan.org/pkg/pifont

\definecolor{viridian}{rgb}{0.25, 0.51, 0.43}
\definecolor{mediumseagreen}{rgb}{0.24, 0.7, 0.44}
\definecolor{otterbrown}{rgb}{0.4, 0.26, 0.13}
\definecolor{saddlebrown}{rgb}{0.55, 0.27, 0.07}
\definecolor{americanrose}{rgb}{1.0, 0.01, 0.24}
\definecolor{ao}{rgb}{0.0, 0.0, 1.0}

\newcommand\myshade{80}
\colorlet{mylinkcolor}{violet}
\colorlet{mycitecolor}{red}
\colorlet{myurlcolor}{ao}

\newcommand{\change}[1]{{\color{black} #1}}

\hypersetup{
	linkcolor  = mylinkcolor!\myshade!black,
	citecolor  = mycitecolor!\myshade!black,
	urlcolor   = myurlcolor!\myshade!black,
	colorlinks = true
}

\usepackage{tikz,xcolor,hyperref}

\definecolor{lime}{HTML}{A6CE39}
\DeclareRobustCommand{\orcidicon}{\hspace{-3mm}
	\begin{tikzpicture}
		\draw[lime, fill=lime] (0,0) 
		circle [radius=0.16] 
		node[white] {\hspace{0.1mm}{\fontfamily{qag}\selectfont \tiny ID}};
		\draw[white, fill=white] (-0.07,0.1) 
		circle [radius=0.01];
	\end{tikzpicture}
	\hspace{-5mm}
}

\foreach \x in {A, ..., Z}{\expandafter\xdef\csname 
	orcid\x\endcsname{\noexpand\href{https://orcid.org/\csname orcidauthor\x\endcsname}
		{\noexpand\orcidicon}}
}

\title{Distinguishing Neutron Star vs. Low-Mass Black Hole Binaries with Late Inspiral \textit{\&} Postmerger Gravitational Waves --- Sensitivity to Transmuted Black Holes and Non-Annihilating Dark Matter}

\author[a]{Sulagna Bhattacharya,\orcidA{}}
\emailAdd{sulagna@theory.tifr.res.in}

\author[b]{Shasvath Kapadia,\orcidB{}}
\emailAdd{shasvath.kapadia@iucaa.in}

\author[a]{and Basudeb Dasgupta\orcidC{}}
\emailAdd{bdasgupta@theory.tifr.res.in}

\affiliation[a\hspace{2pt}]{Tata Institute of Fundamental Research,\\ Homi Bhabha Road, Mumbai 400005, India}
\affiliation[b\hspace{2pt}]{Inter-University Centre for Astronomy and Astrophysics, \\ Post Bag 4, Ganeshkhind, Pune  411007, India}

\begin{abstract}
{\change{Gravitational wave signals from binary neutron star (BNS) mergers and binary low-mass black hole (BLMBH) mergers are highly similar in the early inspiral phase. Consequently, the astrophysical origin of recently detected low-mass compact binary coalescences has remained ambiguous, particularly in the absence of electromagnetic counterparts. In this work, we demonstrate that proposed detectors with increased high-frequency sensitivity -- including NEMO, Cosmic Explorer, and the Einstein Telescope -- will reliably distinguish these two source classes in the late inspiral and postmerger regimes. We further show how these detections can be used to disentangle the individual contributions of BNS and BLMBH systems to the compact binary merger rate, while accounting for misclassification probabilities. Finally, we show this can lead to constraints on the interaction of heavy, non-annihilating dark matter with nucleons. This is achieved by noting that capture of such dark matter particles into neutron stars would lead to transmuted black holes (TBHs), formed via neutron star collapse, which would contribute to the BLMBH rate.}}
\end{abstract}

\subheader{TIFR/TH/25-10}

\begin{document}
	\maketitle
	\flushbottom
	
	\section{Introduction}
	\label{intro} 
 Gravitational wave (GW) \cite{Thorne:1980rt} observations of compact binary mergers provide key insights into strong-field gravity and relativistic astrophysics. While GW150914\,\cite{LIGOScientific:2016aoc}, the first detected coalescence of two black holes (BHs), has confirmed the existence of stellar-mass binary black hole (BBH) mergers and provided a new way to study general relativity, GW170817\,\cite{LIGOScientific:2017vwq}, the first observed merger of two neutron stars (NSs), has opened a new window to multimessenger astronomy by linking gravitational waves to electromagnetic counterparts. During the O3a+b observing runs, the LIGO-Virgo-KAGRA (LVK) Collaboration\,\cite{LIGO_Scientific_Collaboration} has reported a number of compact binary coalescence (CBC) events with component masses that resist clear classification as either NSs or low-mass black holes (LMBHs). Examples include GW190425\,\cite{LIGOScientific:2020aai}, GW190814\,\cite{LIGOScientific:2020zkf}, GW230529\,\cite{LIGOScientific:2024elc}, and SSM200308\,\cite{Prunier:2023uoo}, where at least one of the components is either in the lower mass-gap region, viz. $2.5-5\,M_{\odot}$, or even lighter\,\mbox{\cite{Bailyn:1997xt, Ozel:2010su, Farr:2010tu}}. Such LMBHs are not expected as a product of standard stellar evolution, and thus the current practice is to classify CBC components as NSs or BHs using the inferred mass (i.e. NS if $\lesssim2.5\,M_{\odot}$, and BH if not). However, such a classification is provisional and a more rigorous data-driven approach is warranted.
	
In this paper, we therefore ask the question --- {\change{\emph{How well can we distinguish NS-NS and LMBH-LMBH mergers?}} This is motivated from two complementary viewpoints. Firstly, from an astrophysical/observational standpoint, it is desirable to classify a low-mass CBC as a binary neutron star (BNS) or a binary low-mass black hole (BLMBH) merger, assuming these are the only two possibilities. This is important in order to obtain unbiased inferences on the NS population and merger rates. Secondly, from a fundamental physics perspective, the potential existence of these exotic LMBHs offers a novel gravitational avenue to discover new physics\,\mbox{\cite{Carr:2020xqk, Abramowicz:2022mwb, Dasgupta:2020mqg, Singh:2020wiq, Bhattacharya:2023stq}}. 
	
 A specific new physics possibility involves dark matter (DM)\,\mbox{\cite{Bramante:2023djs, Cirelli:2024ssz}}, which constitutes ~27\% of the Universe’s energy content. Despite its ubiquity, its particle identity remains elusive -- with its mass uncertain within a huge range ($10^{-22}-10^{66}\,\rm eV$)\,\cite{Balazs:2024uyj} and no confirmed non-gravitational interactions. Heavy non-annihilating DM particles are capable of producing LMBHs via DM capture-induced collapse of a NS -- i.e., a `transmuted black hole' (TBH); we discuss this in more detail in Sec.\,\ref{sec:TBH}. Interestingly, mergers of these TBHs can mimic BNS merger signals\,\cite{Bhattacharya:2023stq, Dasgupta:2020mqg}, and their apparent absence places constraints on DM mass and nucleon cross sections\,\cite{Bhattacharya:2023stq, KAGRA:2021duu}. More broadly, GW observations can serve as a novel probe of DM~\cite{Bertone:2019irm}, complementing traditional direct, indirect and collider-based searches in a variety of ways.\footnote{Other related works connecting DM with GW observations include the exploration of DM-admixed BNS mergers~\cite{Giangrandi:2025rko, Karkevandi:2021ygv, Mukherjee:2025omu}, inferring DM-induced modifications to NS oscillation modes~\cite{Shirke:2023ktu, Shirke:2024ymc}, and probing DM density spikes around super-massive black holes~\cite{Tiwari:2025qqx} using extreme mass ratio inspirals (EMRIs)~\cite{Li:2021pxf, Cole:2022ucw, Karydas:2024fcn, Kavanagh:2024lgq}. In addition, GW observations have also shed light on the possible nature of DM by probing dark photons as DM\,\cite{Pierce:2018xmy}, ultralight scalars\,\cite{Morisaki:2018htj}, non-annihilating DM\,\mbox{\cite{Bhattacharya:2023stq, Bhattacharya:2024pmp, Singh:2022wvw}}, compact DM\,\mbox{\cite{Sasaki:2016jop, Bird:2016dcv, LIGOScientific:2019kan, DeLuca:2020sae}}, and self-interacting DM\,\mbox{\cite{Kadota:2023wlm, Han:2023olf, Adhikary:2024btd}}, among others.}
	
Coming back to the issue of distinguishing BNS and BLMBH mergers, a key distinguishing feature between them is the potential observation of an electromagnetic counterpart through multi-messenger detection. As in the case of GW170817\,\cite{LIGOScientific:2017vwq}, non-gravitational signals would provide clear evidence against BLMBHs. However, it is estimated that most of the CBC detections at current and future GW detectors will not be adequately followed up by electromagnetic and neutrino searches\,\mbox{\cite{Saleem:2017pvc, Stratta:2022ufl, Nicholl:2024ttg}}. This has to do with the limited sky coverage and/or angular resolution of the relevant telescopes, as well as imperfect localization of the GW sources. 
	
In the absence of electromagnetic or neutrino counterparts, it is challenging to confidently distinguish a BNS merger from a BLMBH merger. Current GW detectors primarily record the inspiral phase signal, where the waveforms of these systems are nearly identical. A promising approach involves measuring tidal effects during the inspiral --- Neutron stars, being extended objects with internal structure, exhibit non-zero tidal deformability; whereas black holes, being point-like singularities without internal structure, have zero tidal polarizability. Accurate inference of the tidal parameters in the waveform during the inspiral phase thus offers a potential diagnostic to distinguish BNS from BLMBH mergers\,\cite{Singh:2022wvw, Abac:2023ujg}. Several recent studies have explored this avenue\,\mbox{\cite{Fishbach:2020ryj, Essick:2020ghc, Datta:2020gem, Farah:2021qom, Mukherjee:2022wws, Mukherjee:2023pge, Crescimbeni:2024qrq, DeLuca:2024uju, Crescimbeni:2024cwh, Golomb:2024mmt}}. \change{However, it is important to note that tidal effects can be degenerate with other effects such as spin or orbital eccentricity, unless high-order post-Newtonian effects are reliably modeled and detected\,\cite{Favata:2013rwa}. A reliable extraction of tidal signatures demands favorable observational conditions: nearby sources, long inspiral durations with high signal-to-noise ratios (SNRs), and enhanced detector sensitivity at the higher end of currently detectable frequencies (note that the merger frequencies for $1\,M_{\odot}$-$1\,M_{\odot}$ systems are approximately 2\,kHz)\,\cite{Maggiore:2007ulw}. In contrast, matter effects in neutron stars become more pronounced and qualitatively distinct in the postmerger phase \cite{Lattimer:2000nx,Burgio:2021vgk,  Abac:2023ujg}.}

To address the question posed at the beginning, we therefore investigate the similarity between similar-mass NS-NS and LMBH–LMBH mergers, by comparing their waveforms in the late inspiral and postmerger epochs. We assume that the true events are in fact BNS mergers, and ask how well such events can be fit by some BLMBH waveform (see Sec.\,\ref{sec:FF}). This is then statistically quantified using fitting factors and Bayes factors, and we predict the probability to misclassify a true BNS merger as a BLMBH merger (see Sec.\,\ref{sec:BF}).

The main outcome of these studies is an estimate of the probability to correctly classify a CBC as a BNS or a BLMBH. The results depend on the equation of state (EoS) of the neutron star matter and on the detector sensitivity. We then use this to estimate the most probable BNS and BLMBH contributions to the observed CBC rate, as well as to obtain exclusion sensitivity for the fraction of CBCs that could be BLMBHs (see Sec.\,\ref{sec:LMBHfrac}).  Finally, we explore a physically motivated model for formation of LMBHs, i.e. via transmutation of NSs into similar mass BHs due to capture of heavy non-annihilating DM particles, and compute exclusion sensitivity to the DM mass $m_\chi$ and its interaction cross section $\sigma_{\chi n}$ with nucleons (see Sec.\,\ref{sec:TBH}). 
 
 \begin{figure}[t]
	\centering
	\includegraphics[width=0.8\textwidth]{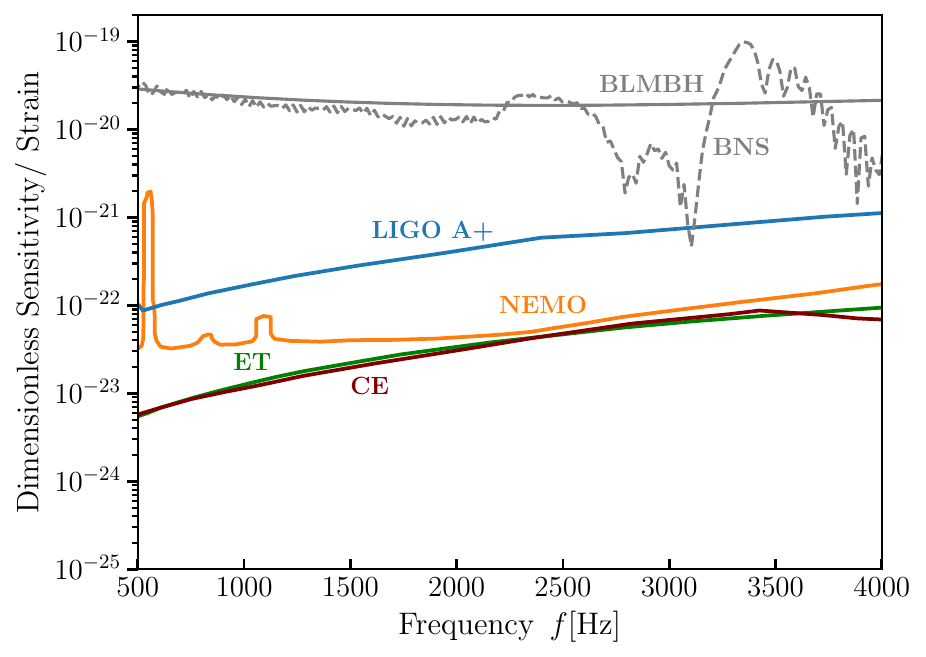}
	\caption{Dimensionless strain amplitudes $f\tilde{h}(f)$ for a benchmark BNS merger and BLMBH merger (grey lines), and the amplitude spectral density (ASD) of the detector noise for the set of current and proposed detectors, LIGO A+\,\cite{LIGOScientific:2014pky}, NEMO\,\cite{Ackley:2020atn}, CE\,\cite{Reitze:2019iox}, and ET\,\cite{Punturo:2010zz}, considered in this study. Here the source is taken at a luminosity distance $D_L=1 $ Mpc. The different ASDs are shown as $\sqrt{fS_n(f)}$ to keep it dimensionless. $S_n(f)$ denotes the one-sided power spectral density of the detector noise.}
	\label{fig_Snvsf}
\end{figure}
 
\section{Distinguishability of BNS \emph{\&} BLMBH Mergers}
\label{sec:methods}

In Fig.\,\ref{fig_Snvsf} we show the dimensionless strains for example symmetric BNS and BLMBH mergers, in comparison to the noise sensitivity curve of each of the detectors: \change{LIGO A+\,\cite{LIGOScientific:2014pky}, NEMO\,\cite{Ackley:2020atn}, CE\,\cite{Reitze:2019iox}, and ET\,\cite{Punturo:2010zz}}. See Appendix \ref{sec:A2} for details of each of these detectors. The key observation is that, while at lower frequencies ($\lesssim 2\,$kHz) the waveforms are qualitatively similar (though quantitatively different), at higher frequencies the BNS merger produces a distinctive second peak and the signals have qualitative differences.

\change{To quantify the distinguishability of the BNS and BLMBH waveforms,  we will treat the BNS waveform for a given EoS as the putative detected waveform and fit it with BLMBH merger waveforms. The overlap between two waveforms is termed as match~$\mathcal{M}$. The maximal match, obtained by scanning over the intrinsic parameters of BLMBH templates, is termed as the fitting factor $\rm FF$, which can then be used to estimate a Bayes factor BF. In the following, we first list the benchmark parameters and EoSs for our study, our procedure for calculating the FFs and BFs, and then compute the detector-specific distinguishing ability for the chosen set of EoSs.}

\newpage

\subsection {Inputs \emph{\&} Benchmark Parameters}
\label{sec: benchmark}

\change{As our benchmark BNS waveforms, we use a set of 8 numerical relativity (NR) waveforms from the CoRe database\,\cite{CoreWebsite}. Of these, 6 are BAM waveforms with zero-temperature EoSs  (2H, MS1b, H4, ALF2, SLy, 2B) and 2 are finite-temperature THC EoSs ($\rm BHB\Lambda\phi$ and LS220). See Appendix\,\ref{sec: BNS waveform} for details of the CoRe database, EoSs, and practicalities of extracting the waveforms. All of these correspond to a merger of two non-spinning NSs, each with mass $1.35\,M_\odot$, in a non-eccentric orbit.}

\change{As our template bank, we use BLMBH merger waveforms generated using the IMRPhenomD\,\mbox{\cite{Husa:2015iqa, Khan:2015jqa}} model in PyCBC\,\cite{alex_nitz_2024_10473621}. In this study, we restrict ourselves to equal-mass binaries, with each component mass in the range $1-2.5\,M_\odot$. We also assume that they are non-spinning and the orbits are not eccentric. See Appendix\,\ref{sec:LMBH waveform} for further details.}

\begin{table}[t]
	\centering
	
	%================= Left panel (a) =================
	\begin{minipage}[t]{0.48\linewidth}
		\centering
		{(a) Intrinsic Parameters}\\[3pt]
		\scalebox{0.95}{
			\begin{tabular}{ccc}
				\toprule 
						& \textbf{BNS} & \textbf{BLMBH}\\
				\midrule
				 $m_1=m_2$ & $1.35\,M_\odot$ & $1 - 2.5\,M_\odot$\\
				$\chi_1=\chi_2$ & 0 & 0\\
				Microphysics & See Table\,\ref{tab:fmerge} & --- \\
				\bottomrule
			\end{tabular}}
	\end{minipage}%
	\hspace{0.5 em}
	%================= Right panel (b) =================
	\begin{minipage}[t]{0.48\linewidth}
		\centering
		{(b) Distances}\\[3pt]
	\scalebox{0.95}{
			\begin{tabular}{lc}
				\toprule
				\textbf{Detector} & {$D_L$ [Mpc]} \\
				\midrule
				LIGO A+   & 100 \\
				NEMO      & 300 \\
				CE \emph{\&} ET  & 350 \\
				\bottomrule 
			\end{tabular}}
	\end{minipage}
	\caption{Benchmark intrinsic parameters, component masses ($m_1,\,m_2$) \emph{\&} dimensionless spins ($\chi_1,\,\chi_2$), used in this study for BNS and BLMBH systems. All sources are assumed to be equal-mass, non-spinning binaries. The benchmark source luminosity distance $D_L$ for each detector analyzed in this work is listed alongside.}
\label{tab:benchmark_params}

\end{table}

\begin{table}[h]
	\centering
	{BNS Microphysics}\\[3pt]
	\scalebox{1}{%
		\begin{tabular}{c   c   c   c   c  }
			\toprule
			\textbf{EoS} & \textbf{CoRe ID} & ${\kappa_2^T}$&  $\tilde{\Lambda}$   &$f_{\rm merge}$ [Hz] \\ 
			\midrule
			{2H} & BAM:0002&436&2325& 1230\\ 
			{MS1b}& BAM:0065 &287&1531 & 1407 \\ 
			{H4}& BAM:0035 &208&1111& 1554\\
			{BHB$\Lambda\phi$} & THC:0003&159 &848& 1677 \\ 
			{ALF2} & BAM:0003&137&733& 1743\\ 
			{LS220} & THC:0019&128& 684&1774\\ 
			{SLy} & BAM:0098&73&390 &2001\\ 
			{2B} &BAM:0001&24&127&2298\\ 
			\bottomrule
		\end{tabular}
	}
		
	\caption{The EoSs, considered for \change{the BNS systems in} this study, with their CoRe IDs. The reduced tidal parameters $\kappa_2^T$ \emph{\&} $\tilde{\Lambda}$ are also listed (see Eq.\,\ref{eq:eos} for definitions). The last column gives the predicted merger frequencies\,\mbox{\cite{Breschi:2019srl, Vijaykumar:2022fst, Bernuzzi:2014kca}}. We take the frequency range from 500\,Hz to $f_{\rm merge}$ as the inspiral and from $f_{\rm merge}$ to 4000\,Hz as the postmerger.}
	\label{tab:fmerge}
\end{table}

Table~\ref{tab:benchmark_params} lists the intrinsic parameters used for both BNS an BLMBH waveforms. \change{It also notes the benchmark distances $ D_L $} that we have assumed for sources to be detected at each of the GW detector considered. These benchmark distances are loosely chosen to ensure a sufficiently significant detection in each case, and we show the dependence of our results on $D_L$ in Sec.\,\ref{sec:DL}. We also list \change{the EoSs used in this study} in Table~\ref{tab:fmerge} from stiffer to softer models\,\mbox{\cite{Breschi:2019srl, Gonzalez:2022mgo}}. 

\subsection{Fitting Factor}
\label{sec:FF}
The detector signal is modeled as a time-series, $s(t)=h(t)+n(t)$, where $h(t)$ is the signal strain and $n(t)$ is the noise in the detector, which is match-filtered with a bank of templates $h_{T}(t)$ to generate the signal to noise ratio (SNR) statistics of a particular event\,\mbox{\cite{Wainstein:1970,Creighton:2011zz, Maggiore:2007ulw}}. The optimal SNR is obtained when $h_T(t)=h(t)$, and given as
\begin{align}
	\label{optsnr}
	{\rm SNR_{\rm opt}}\equiv\rho_{\rm opt}= \sqrt{4\int_{f_{\rm min}}^{f_{\rm max}}\frac{|\tilde{h}(f)|^2}{S_n(f)}df}\equiv\,{\rm norm}(h)\,,
\end{align}
where $\tilde{h}(f)$ is the Fourier transform (frequency domain) of the  time domain data $h(t)$ and $S_n(f)$ is the one-sided power spectral density (PSD) of the detector\,\cite{Maggiore:2007ulw},
\begin{align}
	\langle\tilde{n}(f)\tilde{n}^*(f^{'})\rangle=\frac{1}{2}\delta(f-f^{'})S_n(f)\,.
\end{align}

In reality, there remains some mismatch between the template $h_T$ and the detector data, which can be further analyzed to infer the parameter values and errors associated with each of them.  If the noise is stationary and Gaussian, the likelihood is\,\cite{Cutler:1994ys}
\begin{align}
	\label{likelihood}
	\mathcal{L}= p(s|\vec{\theta})\propto \, e^{-\frac{1}{2}\langle s-h_T(\theta)|s-h_T(\theta)\rangle}\,,
\end{align}
where $\vec{\theta}$ are the model parameters with which the template has been constructed, and $s$ is the signal received. The notation $\langle.|.\rangle$ denotes the noise weighted inner product given by\,\mbox{\cite{Cutler:1994ys, Finn:1992wt}}
\begin{align}
	\label{inner product}
	\langle h_1|h_2 \rangle=2\int_{f_{\rm min}}^{f_{\rm max}} df \frac{\tilde{h}_1^*(f)\tilde{h}_2(f) + \tilde{h}_2^*(f)\tilde{h}_1(f)}{S_n(f)} \,,
\end{align}
with $f_{\rm min}\, \emph{\&} \,f_{\rm max}$ being the minimum and maximum of the frequency over which the inner product is evaluated. This frequency range is discussed later in the context of the problem's requirement.

Our aim is to quantify \change{how well a BLMBH merger waveform matches with a given BNS merger waveform}. This comparison can be performed by calculating the fitting factor FF between two waveforms $h_1$ (signal = BNS) and $h_2$ (template = BLMBH),
\begin{align}
	\label{eq:FF^1_2}
	\nonumber
	&{\rm FF} = \max_{\text{templates}} \mathcal{M} =\max_{\text{templates}} 2 \int_{f_{\rm min}}^{f_{\rm max}} df \frac{\left(\tilde{h}_1^*(f)\tilde{h}_2(f)+ \tilde{h}_1(f)\tilde{h}_2^*(f)\right)}{S_n(f)\times \sqrt{\langle {h}_1|{h}_1\rangle}\times \sqrt{\langle {h}_2|{h}_2\rangle}}\,.
\end{align}
This is the match, $\mathcal{M} =  \langle \hat{h}_1(f)|\hat{h}_2(f)\rangle$, maximized over intrinsic parameters of the models\,\mbox{\cite{Finn:1992wt, Cornish:2011ys}}, and used to estimate the maximum overlap achievable between an observed signal (say $h_1$) and the best-matching waveform within a given template bank (say $h_2$). It is normalized, which makes it independent of the extrinsic parameters which affect the norms (for our case, mainly $D_L$). The $\rm FF$ of exactly the same waveforms is unity, and any deviation of the template from the signal will lead to smaller $\rm FF$ values, resulting in the SNR loss in the detector,
\begin{align}
	\text{SNR}_{\rm observed} = \rm FF \times \text{SNR}_{\rm opt}\,.
\end{align}
A reduction in $\rm FF$ by more than 0.03 corresponds to a $\gtrsim$ 10\% loss in the recovered signal amplitude, and hence leads to a significant decrease in the evidential support for the concerned waveform\,\mbox{\cite{Owen:1995tm, Dhurandhar:1992mw}}.

For our analysis we consider the frequency range from $500-4000$\,Hz, where we call the range $500\,{\rm Hz} - f_{\rm merge}$(depending on $\kappa_2^T$, hence EoS) as the late inspiral phase and $f_{\rm merge}-4000$\,Hz as the postmerger phase. The sensitivity drops after 4000\,Hz, resulting in no significant contribution from higher frequencies (Fig.\,\ref{fig_Snvsf}). We use an analytical fit from Ref.\,\cite{Breschi:2019srl} to evaluate $f_{\rm merge}$ using
\begin{align}
	Mf_{\rm merge}= \left(3.3184\times 10^{-2}\right)\frac{1+\left(1.3067\times10^{-3}\right)\zeta_{3199.8}}{1+\left(5.0064\times10^{-3}\right)\zeta_{3199.8}}\,,
\end{align}
where $\zeta_{3199.8} = \kappa_2^T + 3199.8 \left(1-4\nu\right)$, with $\nu$ being the symmetric mass ratio (see Appendix \,\ref{sec: BNS waveform} for further details). 

\begin{figure}[!thb]
\vspace{-0.5cm}
	\centering
	\hfill \quad\quad Inspiral\qquad\qquad \hfill \quad Postmerger\qquad \qquad\qquad \hfill Total \qquad \qquad\qquad\hfill \\
	\includegraphics[width=0.325\textwidth]{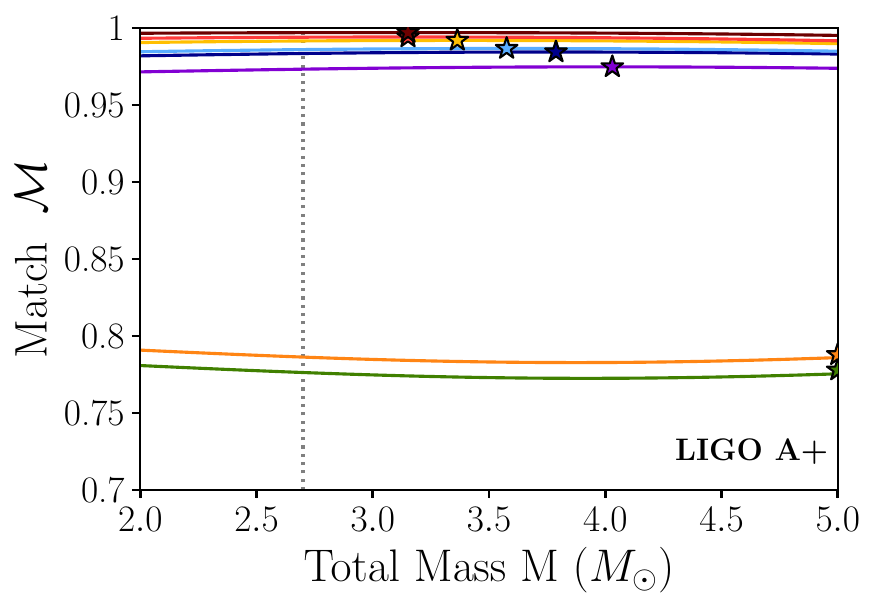}
	\includegraphics[width=0.325\textwidth]{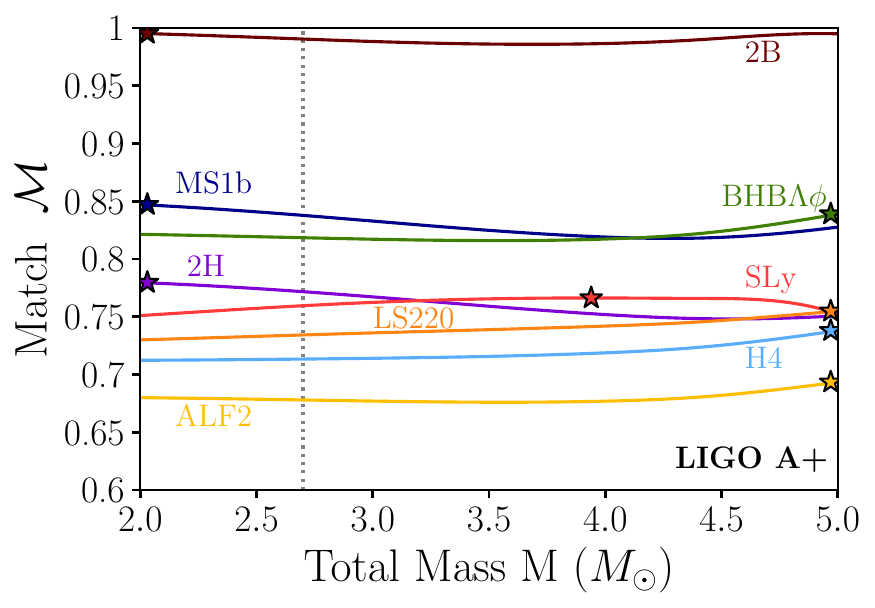}
	\includegraphics[width=0.32\textwidth]{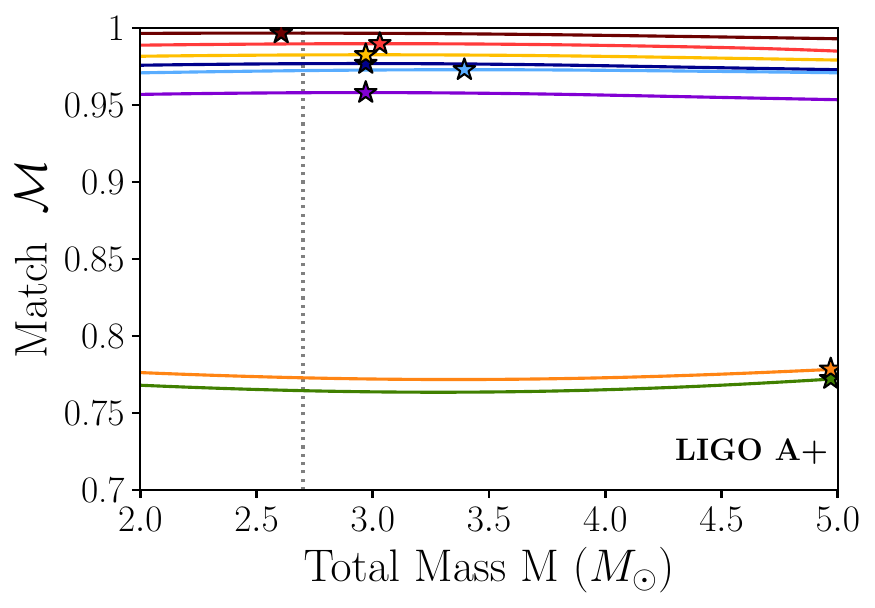}
	
	\includegraphics[width=0.325\textwidth]{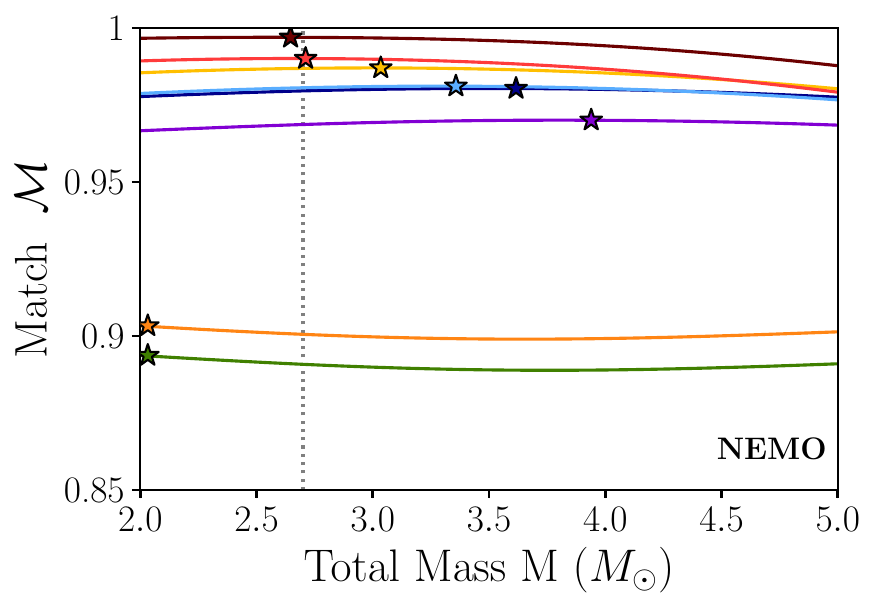}
	\includegraphics[width=0.325\textwidth]{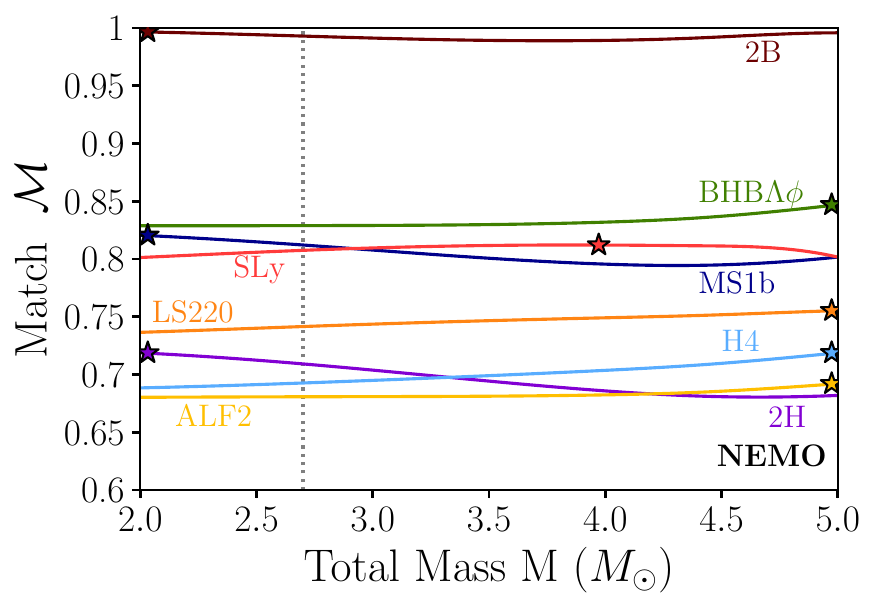}
	\includegraphics[width=0.32\textwidth]{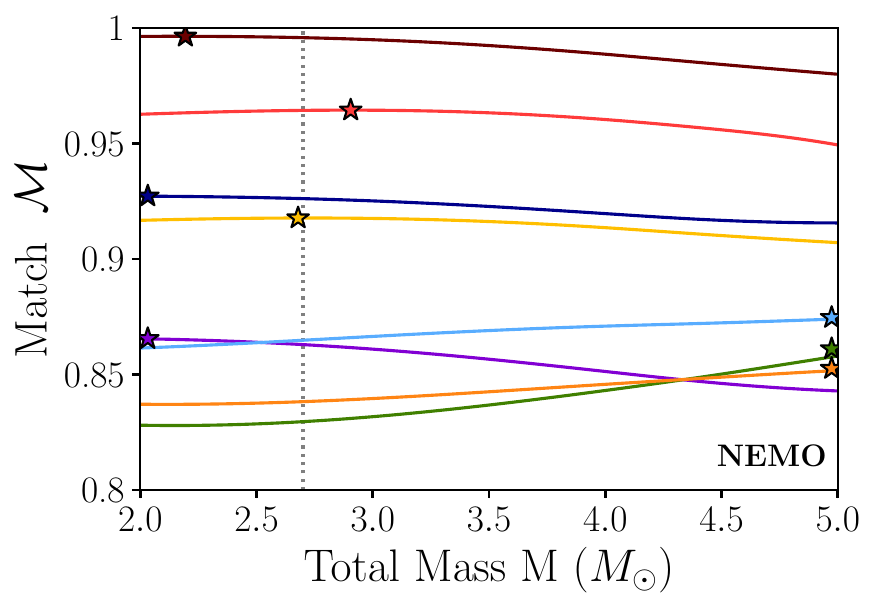}
	
	\includegraphics[width=0.325\textwidth]{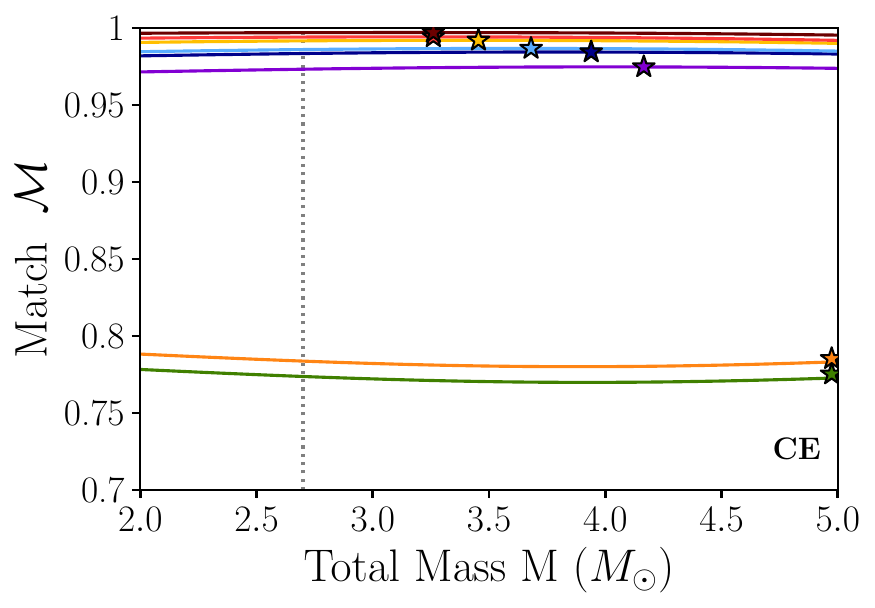}
	\includegraphics[width=0.325\textwidth]{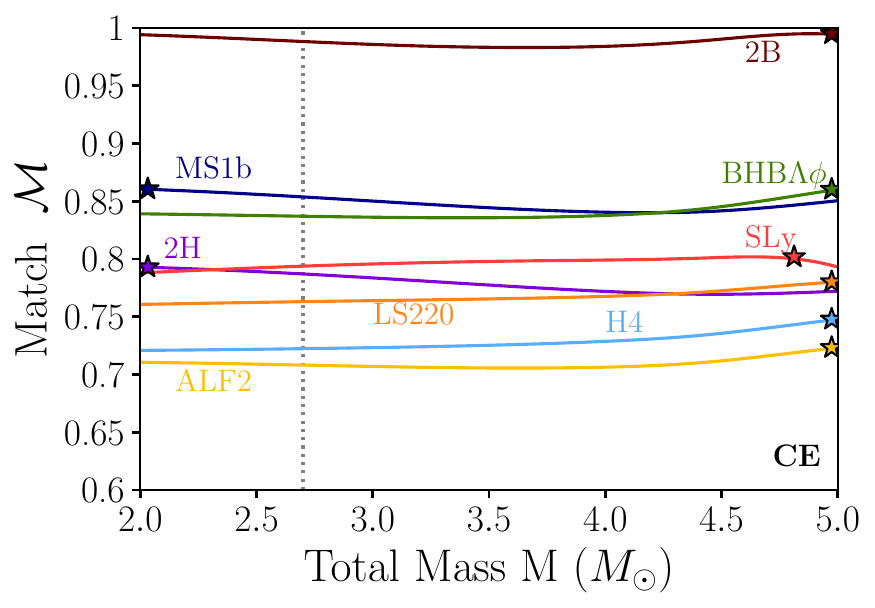}
	\includegraphics[width=0.32\textwidth]{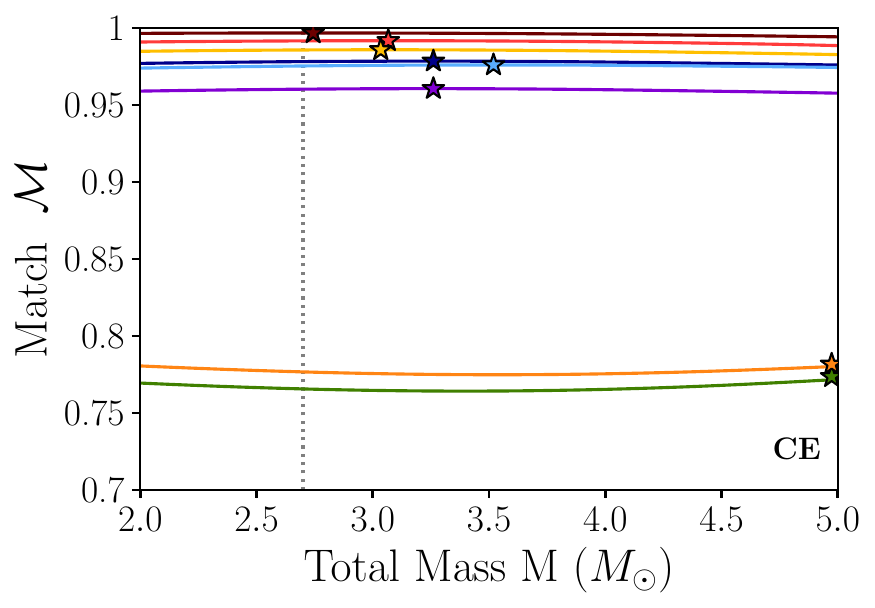}
	
	\includegraphics[width=0.325\textwidth]{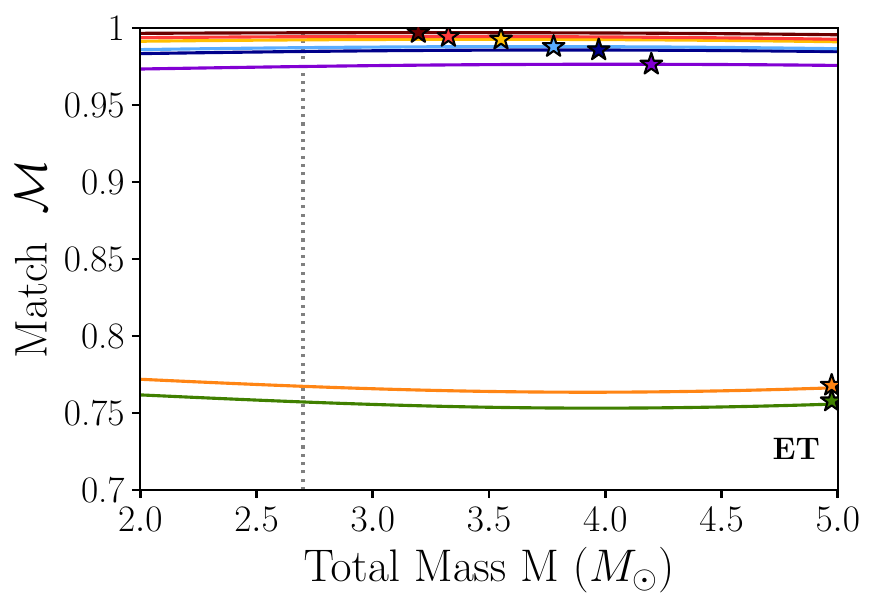}
	\includegraphics[width=0.325\textwidth]{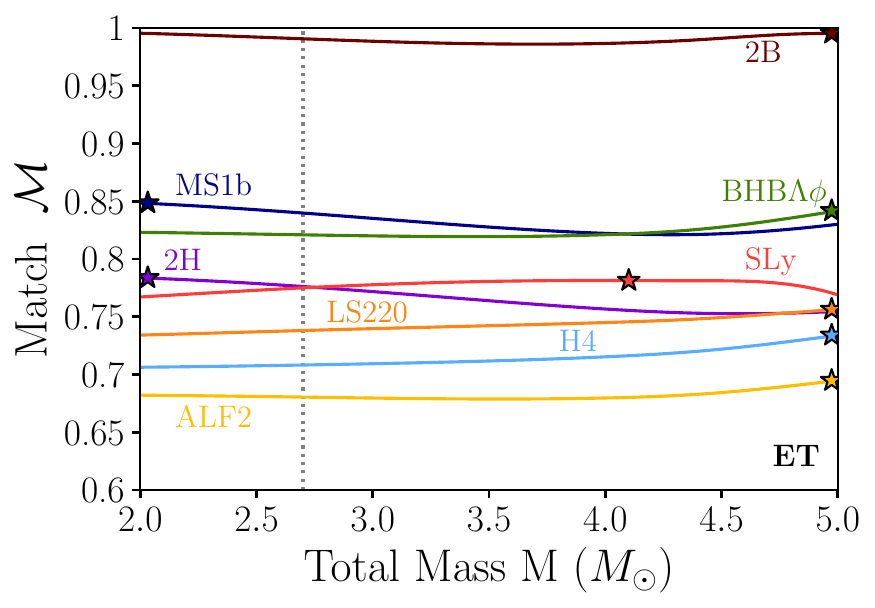}
	\includegraphics[width=0.32\textwidth]{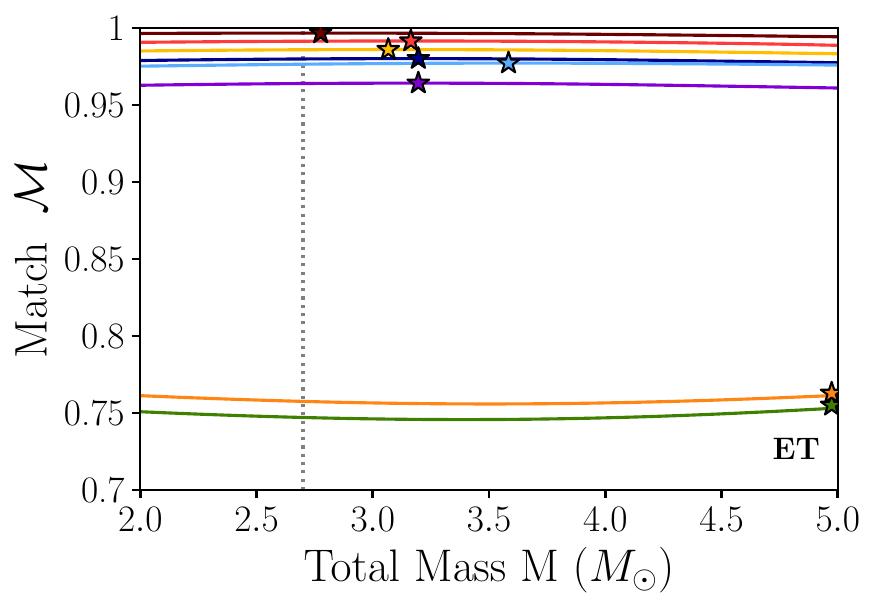}
\caption{\change{The match $\mathcal{M}$ between BNS \emph{\&} BLMBH mergers is shown as a function of the total mass $M=2m_1=2m_2$. Each row corresponds to a different detector, LIGO A+, NEMO, CE, and ET, while the columns correspond to the inspiral, postmerger, and the total signal. The different colored lines correspond to different EoSs (Table\,\ref{tab:fmerge}), with the legend shown in the middle panels. The maximal match is taken as the fitting factor in each case (denoted by a star);  the vertical dotted line denotes the case $m_1 = m_2 = 1.35\,M_{\odot}$, for comparison. The fitting factor in the postmerger signal (middle column) is generally poorer, implying its potentially greater discriminating ability. However, the total FF (right panels) is more similar to that in the inspiral phase because of its greater SNR -- except for NEMO. Notably, two EoSs, BHB$\Lambda\phi$ \emph{\&} LS220, are strongly mismatched in both inspiral and postmerger phases.}}
	\label{fig_match_ff}
\end{figure}

Since analytically varying all physical parameters for BNS systems is impractical due to the simulation costs, we restrict our analysis to eight representative EoSs given in Table~\ref{tab:fmerge}. Table~\ref{tab:fmerge} shows the $\kappa_2^T$ values and predicted merger frequencies.

In Fig.\,\ref{fig_match_ff} we show the matches and fitting factors for our chosen suite of EoSs and detectors. One can see that the fitting factor is generically poor ($<0.97$, corresponding to a loss $>10\%$ of the SNR) in the postmerger signal. However, the norm (or equivalently SNR) is usually larger in the inspiral phase (except for NEMO), and the fitting factor for the total signal more closely matches what would have been obtained from the inspiral signal alone. This illustrates the value of instruments that are capable of detecting the high-frequency postmerger signal to distinguish BNSs and BLMBHs. \change{ The key point is that the matter effect, encoded via the EoS, determines a characteristic secondary peak in the postmerger GW waveform, that in principle provides a clear and robust distinction between the two cases.}\, \mbox{\cite{Radice:2018pdn, Bauswein:2013jpa, Hotokezaka:2011dh}}. As we shall see shortly, this is unfortunately difficult to leverage because of the relatively lower sensitivity at higher frequencies for all detectors, except at NEMO.

\subsection{Bayes Factor}
\label{sec:BF}

The Bayes factor (BF) is the ratio of two marginal likelihoods, which determines whether hypothesis 1 ($\mathcal{H}_1$ = BNS) is favored over hypothesis\,2 ($\mathcal{H}_2$=BLMBH). More accurately, $\mathcal{B}^1_2$ is the evidence of $\mathcal{H}_1$ over $\mathcal{H}_2$, given the signal $s = h_{\rm BNS}$. In our case, the functional form of the BF can be approximated (using a saddle-point argument) in terms of the FF between the two models (with equal a priori odds) and the optimal SNR $\rho_{\rm opt}$ (Eq.\,\ref{optsnr})\,\mbox{\cite{Cornish:2011ys, Vijaykumar:2022fst}}, as
\begin{align}
	\label{eq:Bayes Factor}
	\mathcal{B}^1_2 = \frac{p(\mathcal{H}_1\,|\,s)}{p(\mathcal{H}_2\,|\,s)} = e^{(1-{\rm FF}^2)\frac{\rho_{1\,\mathrm{opt}}^2}{2}}\,.
\end{align}
We interpret $\mathcal{B}^{\rm BNS}_{\rm BLMBH} \equiv \mathcal{B}$ as the relative likelihood of BNS vs. BLMBH models. We will use Eq.\,\ref{eq:Bayes Factor} to estimate $\mathcal{B}$  --- in the late inspiral signal, in the postmerger signal, and in the total signal --- at each detector, and for our suite of EoS.

\subsubsection{Distinguishability at Current \emph{\&} Proposed Detectors}
\subsubsection*{Advanced LIGO}
\begin{table}[h]
	\centering
Distinguishability at LIGO A+ \\[0.5ex]
	\begin{tabular}{c c c c c c c c c}
		\toprule
		\textbf{EoS } & norm (Ins)& norm (PM) & \textbf{$\rm FF_{Ins}$} &  \textbf{$\rm FF_{PM}$} &  \textbf{$\rm FF_{total}$} & $\rm \mathcal{B}_{Ins}$ & $\rm \mathcal{B}_{PM}$ & $\rm \mathcal{B}_{total}$ \\ 
		\midrule
		\textbf{2H} &3.51 & 1.04 & 0.97 & 0.77 & 0.96 & 1.38 & 1.24& 1.73\\ 
		\textbf{MS1b} &3.72 & 0.82 & 0.98 & 0.84 & 0.98 & 1.26 & 1.11 & 1.4\\ 
		\textbf{H4} & 3.89& 0.97 & 0.99 & 0.71 &0.97 & 1.23  & 1.26 &1.55 \\ 
		\textbf{BHB$\Lambda\phi$} &2.31 & 0.87& 0.78 & 0.82 & 0.76 & 2.9 & 1.12 & 3.5 \\ 
		\textbf{ALF2} &4.00& 0.75& 0.99& 0.68 &0.98 & 1.14 & 1.16 &1.33 \\ 
		\textbf{LS220}& 2.35 & 0.76& 0.79&0.73&0.77 &2.89& 1.14 & 3.42 \\ 
		\textbf{SLy} & 4.23 & 0.62 & 0.99 & 0.76 & 0.99 & 1.11 & 1.08& 1.21\\
		\textbf{2B} &4.37& 0.31& 0.997 & 0.99 & 0.997 & 1.06 & 1.00& 1.07\\
		\bottomrule
	\end{tabular}
	\caption{Fitting factor FF \emph{\&} Bayes factor $\mathcal{B}$ for BNS vs BLMBH merger models, for different EoSs, computed separately in the late inspiral phase, the postmerger phase, and in total. The norm, i.e.  $\rho_{\rm opt}$, is shown for the late inspiral and postmerger phases separately; the total norm is obtained by adding these in quadrature. As one can see, despite potential distinguishability due to low FFs in some cases, the BFs are not significant and the BNS and BLMBH models are essentially indistinguishable. $D_L$ is chosen to be 100\,Mpc.}
	\label{tableligo}
\end{table}

In Table~\ref{tableligo} we see that although the FFs are significantly lower than unity in the postmerger phase, and thus potentially discriminating, the BFs are quite close to unity and there is practically no preference for BNS over BLMBH or vice versa. Even for the EoSs  BHB$\Lambda\phi$ \emph{\&} LS220, which already have low FFs in the inspiral phase, the BFs are barely worth a mention. This is simply because the $\rho_{\rm opt}$ at LIGO A+ is not sufficiently high to turn the mismatch into a sizable evidence for either model. Here we chose a benchmark source distance $D_L=100$\,Mpc.

\subsubsection*{NEMO}
\begin{table}[h]
	\centering
	Distinguishability at NEMO \\[0.5ex]
	\begin{tabular}{ c c c c c c c c c}
		\toprule
		\textbf{EoS } & norm (Ins)& norm (PM) & \textbf{$\rm FF_{Ins}$} &  \textbf{$\rm FF_{PM}$} &  \textbf{$\rm FF_{total}$} & $\rm \mathcal{B}_{Ins}$ & $\rm \mathcal{B}_{PM}$ & $\rm \mathcal{B}_{total}$ \\
		\midrule
		\textbf{2H} &3.06 & 2.72 & 0.97 & 0.71 & 0.87 & 1.33 & 6.32 & 8.5 \\ 
		\textbf{MS1b} &3.42& 2.45 & 0.98 & 0.81 & 0.93 & 1.27 & 2.78 & 3.53 \\ 
		\textbf{H4} & 3.67& 3.22& 0.98& 0.7&0.87 & 1.3 & 14.8 & 20.2\\ 
		\textbf{BHB$\Lambda\phi$} &2.95 & 2.72 & 0.89 & 0.83& 0.85 & 2.46 & 3.2 & 12.37 \\ 
		\textbf{ALF2} &3.94 & 2.32 & 0.99 & 0.68 &0.92 & 1.22& 4.23& 5.2\\ 
		\textbf{LS220}& 3.05& 2.26 &0.9 & 0.74 &0.84&2.4&3.15&8.5\\ 
		\textbf{SLy} & 4.32& 1.84 & 0.99 & 0.81 & 0.97 & 1.2 & 1.8 & 2.2 \\ 
		\textbf{2B} &4.71 & 0.98 & 0.997 & 0.99 & 0.997 & 1.07 & 1.01 & 1.11 \\ 
		\bottomrule
	\end{tabular}
	\caption{Fitting factor FF \emph{\&} Bayes factor $\mathcal{B}$ for BNS vs. BLMBH models at NEMO. For the stiffer EoSs, there is potential for distinguishing BNSs and BLMBHs, with a significant role played by the postmerger signal. $D_L$ is chosen to be 300\,Mpc.}
	\label{tableNEMO}
\end{table}
\change{NEMO, being more sensitive in the postmerger phase than in the inspiral, is expected to capture matter effects more efficiently.} Table~\ref{tableNEMO} presents the FFs and BFs. It is promising that at NEMO there is significant evidence to distinguish the models, especially for the stiffer EoSs (H4, 2H, ALF2). Notably, the evidence from the postmerger phase is significant. We have taken a larger benchmark distance, $D_L$\,= 300 Mpc compared to LIGO A+.

\subsubsection*{Cosmic Explorer and Einstein Telescope}
In Table~\ref{tableET} we show the FF and $\mathcal{B}$ at the Cosmic Explorer (CE) and Einstein Telescope (ET). Here we chose $D_L$ as 350\,Mpc. As their sensitivities significantly exceed that of LIGO A+ over the entire frequency range, and that of NEMO in the low frequency range, the BFs are much larger. Now the BNS and BLMBH models can be confidently distinguished for all viable EoSs considered (note that the EoS 2B is currently disfavored from NS mass-radius data).  A key observation here is that the Bayesian evidence is already significantly high in the inspiral phase due to their superior low-frequency sensitivity; the inclusion of the postmerger phase enhances the evidence but no longer plays as crucial a role as for NEMO. If further parameter degeneracies (spins, eccentricities) are taken into account it is possible though that the postmerger signal may be more important than is indicated here. We expect this because the postmerger signal is less likely to be degenerate with other parameters.

\begin{table}[t]
	\centering
\medskip
		Distinguishability at Cosmic Explorer \\[0.5ex]
	\scalebox{0.9}{%
		\begin{tabular}{c c c c c c c c c }
			\toprule
			\textbf{EoS } & norm (Ins)& norm (PM) & \textbf{$\rm FF_{Ins}$} &  \textbf{$\rm FF_{PM}$} &  \textbf{$\rm FF_{total}$} & $\rm \mathcal{B}_{Ins}$ & $\rm \mathcal{B}_{PM}$ & $\rm \mathcal{B}_{total}$ \\ 
			\midrule
			\textbf{2H} &14.51 & 4.11 & 0.98 & 0.79 & 0.97 & 260.15 & 24.8 & 7006.4\\ 
			\textbf{MS1b} &15.4& 3.22 & 0.99 & 0.86 & 0.98 & 49.5 & 4.1 & 210 \\ 
			\textbf{H4} & 16.1& 3.55 & 0.99 & 0.72 &0.98 & 36.7 & 20.3 &747.2 \\ 
			\textbf{BHB$\Lambda\phi$} &9.47 & 3.03 & 0.78 & 0.84 & 0.77 & $\mathcal{O}(10^8)$& 4 & $\mathcal{O}(10^9)$\\ 
			\textbf{ALF2} &16.56& 2.56& 0.99& 0.71&0.987& 9.8 & 5.1 & 53.57\\ 
			\textbf{LS220}& 9.65& 2.52& 0.79 &0.76&0.78& $\mathcal{O}(10^8)$&3.75& $\mathcal{O}(10^9)$\\ 
			\textbf{SLy} & 17.45 & 2.04 & 0.994 & 0.79& 0.992 & 6.01& 2.2 & 13.13\\ 
			\textbf{2B} &18.04& 1.05& 0.997& 0.988  &0.997 & 2.69 & 1.01& 2.88\\
			\bottomrule\\[3ex]
		\end{tabular}
	}

	\centering
			Distinguishability at Einstein Telescope \\[0.5ex]
	\scalebox{0.9}{%
		\begin{tabular}{c c c c c c c c c}
			\toprule
			\textbf{EoS } & norm (Ins)& norm (PM) & \textbf{$\rm FF_{Ins}$} &  \textbf{$\rm FF_{PM}$} &  \textbf{$\rm FF_{total}$} & $\rm \mathcal{B}_{Ins}$ & $\rm \mathcal{B}_{PM}$ & $\rm \mathcal{B}_{total}$ \\ 
			\midrule
			\textbf{2H} &14.74 & 3.67 & 0.98 & 0.78& 0.97 & 213.9 & 14.5 & 3525.5 \\ 
			\textbf{MS1b} &15.56 & 2.92 & 0.99 & 0.84 & 0.98 & 39.19& 3.52& 147.7 \\ 
			\textbf{H4} & 16.22& 3.5 & 0.99& 0.72 &0.98 & 28.3 & 21.4 & 604.8\\ 
			\textbf{BHB$\Lambda\phi$} &9.02 & 3 & 0.77 & 0.83 & 0.76 & $\mathcal{O}(10^8)$ & 4.2 & $\mathcal{O}(10^9)$ \\ 
			\textbf{ALF2} &16.6& 2.6& 0.99& 0.69 & 0.986 & 8.4 & 6.1 &53.13\\ 
			\textbf{LS220}& 9.18 & 2.58& 0.78 &0.74 &0.77 & $\mathcal{O}(10^8)$& 4.6 &$\mathcal{O}(10^9)$ \\ 
			\textbf{SLy} & 17.46 & 2.15 & 0.995 & 0.78 & 0.992 & 5.6 & 2.5 & 14.2 \\ 
			\textbf{2B} &18 & 1.08& 0.997 & 0.991  &0.997 & 2.74& 1.01 &2.93 \\ 
			\bottomrule
		\end{tabular}
	}
	\caption{Fitting factor FF \emph{\&}  Bayes factor $\mathcal{B}$ for BNS vs BLMBH models at CE (upper table) and ET (lower table), with the signal generated using a BNS waveform with the stated EoSs. The large values of $\mathcal{B}$ indicate that BNS and BLMBH models can be confidently distinguished. The source $D_L$ is chosen to be 350 Mpc.}
	\label{tableET}
\end{table}

\subsubsection{Dependence on Luminosity Distance}
\label{sec:DL}
\change{We now show how the distinguishing ability varies as a function of the distance of the source $ D_L $. We can readily conclude from Eq.\,\ref{eq:Bayes Factor} that the dependence is given by $\log\mathcal{B}\propto 1/D_L^2$. It follows from the fact that $\rm FF$ is independent of $D_L$ and $\rho_{\rm opt}$ scales as $1/D_{L}$. One can compare the detectors at a common benchmark source distance, if needed, by scaling the BFs as above.} 
\begin{figure}[t]
	\centering
	\includegraphics[width=0.45\textwidth]{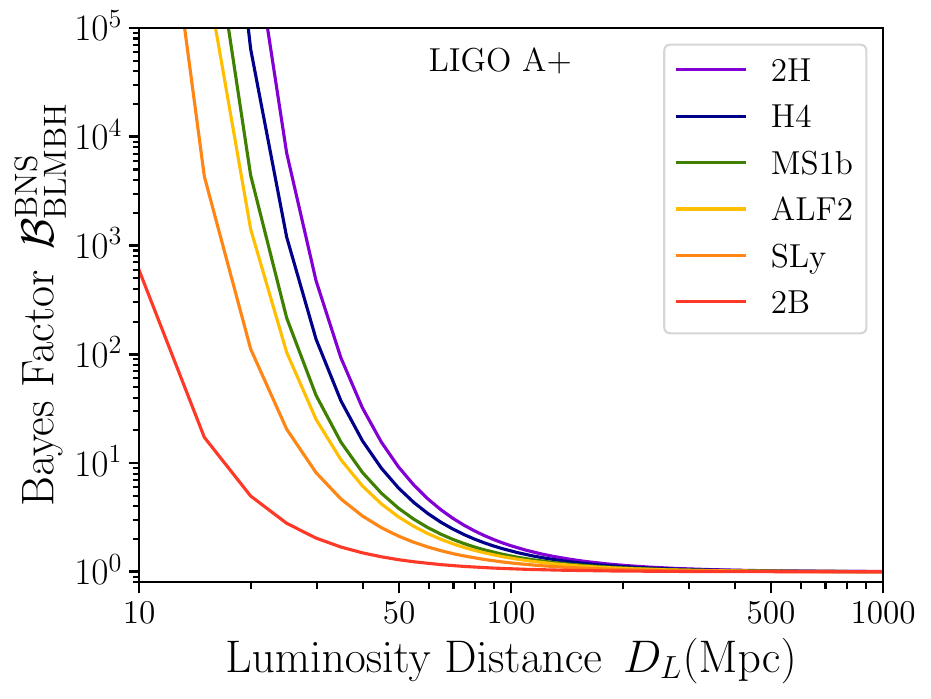}
	\hspace{0.4 cm}
	\includegraphics[width=0.45\textwidth]{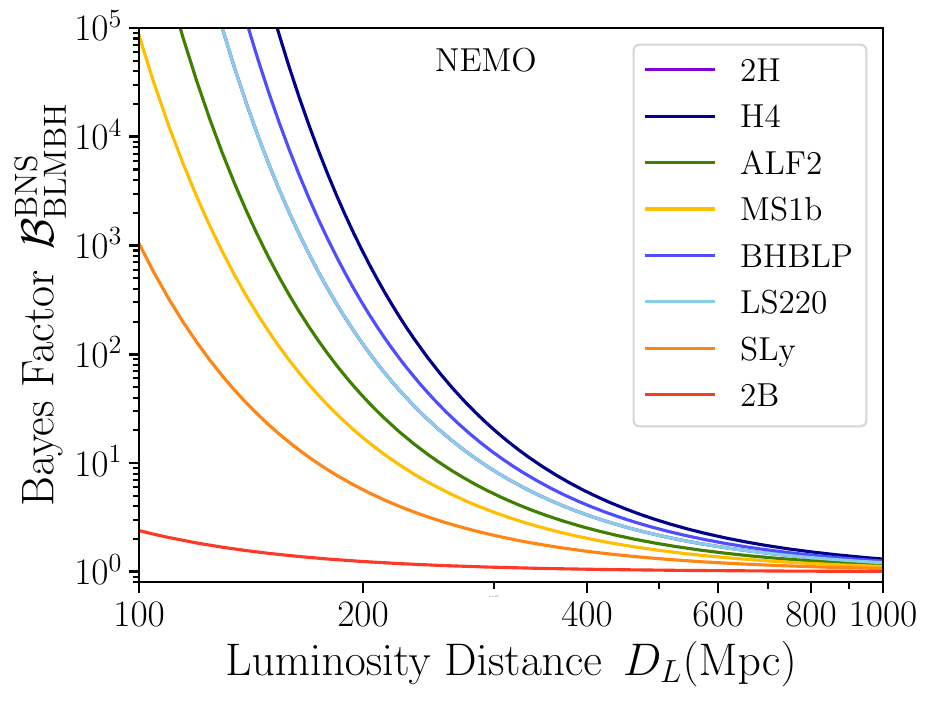} \\[1.5ex]
	\vspace{0.4 cm}
	\includegraphics[width=0.45\textwidth]{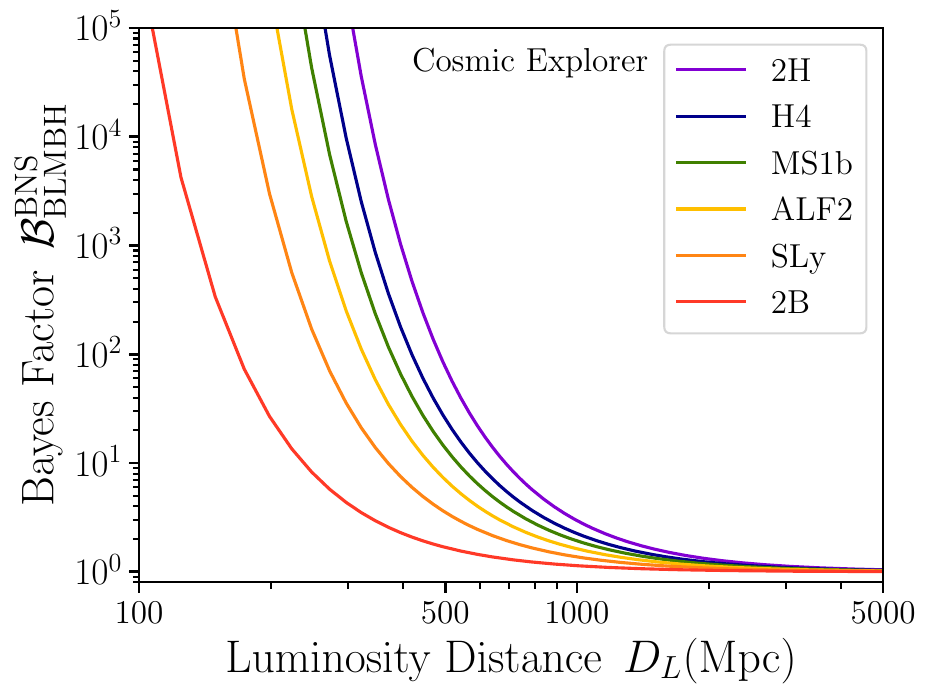}
	\hspace{0.4 cm}
	\includegraphics[width=0.45\textwidth]{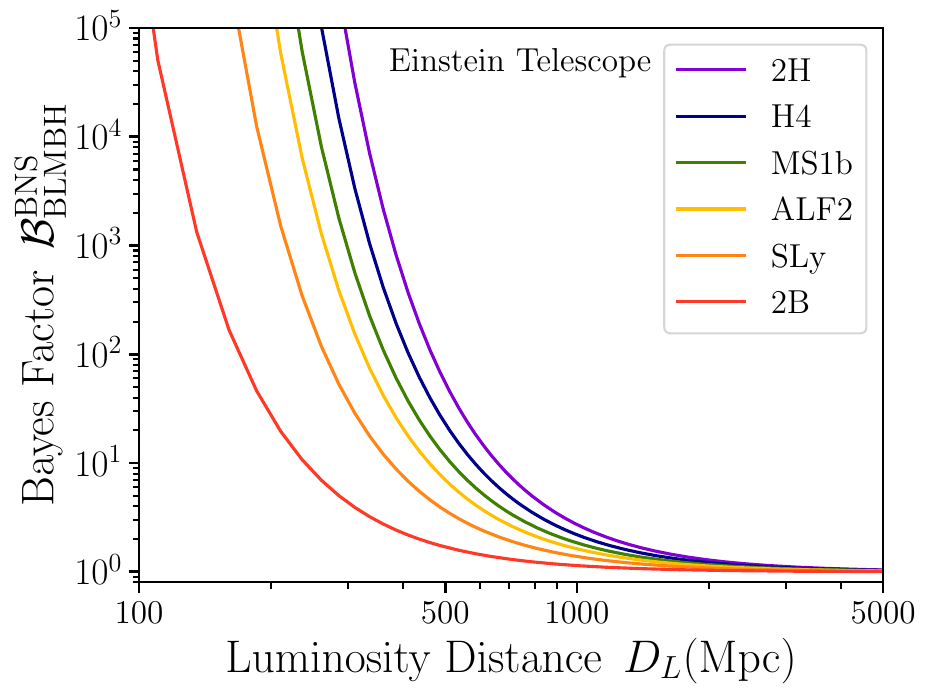}
	\caption{ \change{Bayes Factor ($\mathcal{B}$) as a function of $D_L$ for different detectors and different EoSs. The dependence is given by $\mathcal{B}\propto \exp(1/D_L^2)$. The evidence against a BLMBH merger over a BNS merger becomes significantly large for stiffer EoSs. Note that the distance range on the horizontal axis varies across panels.}}
	\label{fig_BF}
\end{figure} 

\change{In Fig.\,\ref{fig_BF}, we present the total Bayes factor $ \mathcal{B}^{\rm BNS}_{\rm BLMBH} $ evaluated over the full frequency range from 500 Hz to 4000 Hz, as a function of $ D_L $ for the different detectors under consideration.} As before, we consider the BNS merger as the true signal. The strongest Bayesian evidence is observed for stiffer EoS. For the EoSs 2H \emph{\&} H4, the NEMO detector yields a $\mathcal{B}$ exceeding 1000 even at a distance of 200 Mpc, whereas LIGO\,A+ provides negligible evidence at that distance. However, with LIGO\,A+ and with the stiffest EoS 2H, events within 50 Mpc still yield significant evidence. For CE and ET, the evidence remains strong, exceeding a factor of 100,  up to 600 Mpc for the same EoSs. The Bayesian evidence decreases progressively from stiffer to softer EoSs. For the softest EoS (2B), the evidence becomes notably small, as the NSs are sufficiently compact to closely mimic the LMBH waveform. Even with the most sensitive detectors like CE and ET, the Bayes Factor for EoS 2B is only around 11 at 200 Mpc. The intermediate EoSs exhibit evidence values that lie between those of 2H and 2B, and this trend is consistently observed across all detectors.  The two finite-temperature EoSs, $\rm BHB\Lambda\phi$\,\cite{Banik:2014qja} and LS220\,\cite{Lattimer:1991nc},  lead to noticeable differences in the waveform already from the late inspiral phase, resulting in a huge Bayesian evidence from the inspiral phase itself. To be conservative, here we have omitted them.

\section{Constraints on BLMBH \emph{\&} Particle DM }
\label{sec:results}
\subsection{Model-Independent Results on BLMBH Abundance}
\label{sec:LMBHfrac}

\subsubsection{Parsing the CBC Rate into BNS \emph{\&} BLMBH Rates}
\label{RBNS}
As we discussed in the introduction, the current GW data analysis pipeline of the LVK collaboration assigns events with component masses in the range of $1-2.5\,M_{\odot}$ as BNS mergers\,\cite{KAGRA:2021duu}. However, without a simultaneous EM counterpart, such as a kilonova or a short gamma-ray burst, this classification is not robust. This uncertainty becomes critical when estimating the astrophysical merger rate of BNS systems, as it opens up the possibility that some of these observed low-mass events may, in fact, originate from LMBH–LMBH systems, e.g. from primordial BHs or other exotic compact objects. Consequently, it is important to account for this potential contamination to avoid overestimating the true BNS merger rate.

The theoretical estimate of the CBC rate based on stellar evolution and binary formation is given as\,\mbox{\cite{Taylor:2012db, OShaughnessy:2009szr, OShaughnessy:2007brt, deFreitasPacheco:1997fr}}
\begin{align}
	R_{\rm CBC}(t)=\int_{t_*}^{t}dt_f\,\frac{dP_m}{dt}\,\lambda\,\frac{d\rho_*}{dt_f}\,.
	\label{eq:BNSrate}
\end{align}
Here $t_f$ denotes the binary formation time, and ${d\rho_*}/{dt_f}$ denotes the cosmic star formation rate at time $t_f$\,\cite{Madau:2014bja}. The fraction of stellar mass that got bound as a binary is given as $\lambda\,=\,10^{-5}M_{\odot}^{-1}$\,\cite{Taylor:2012db}. The merger time delay distribution ${dP_m}/{dt}\propto 1/(t-t_f)$ is the probability density that the binary formed at time $t_f$ will merge within a time $t-t_f$\,\cite{Taylor:2012db}. The time $t_*=4.9\times10^8$ years corresponds to $z=10$, and is taken as the epoch of formation of the first stars. 

\begin{figure}[t]
	\centering
	\includegraphics[width=0.435\textwidth]{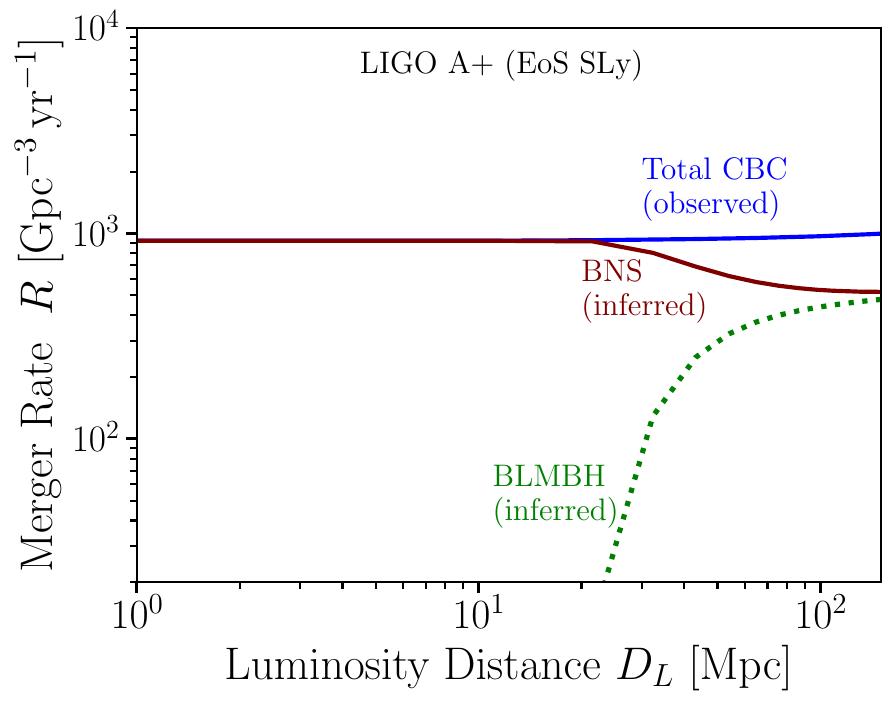}
	\hspace{0.7 cm}
	\includegraphics[width=0.435\textwidth]{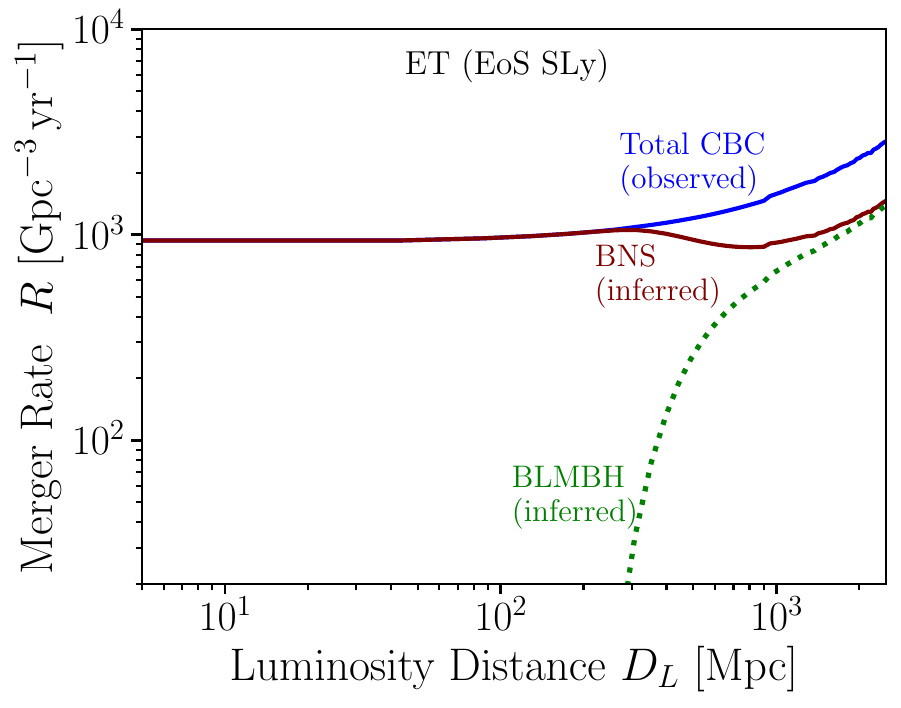}
	\caption{Merger rates as a function of $D_L$ observable by LIGO\,A+\,(left panel) and ET (right panel). The blue solid lines show the theoretically predicted redshift dependence of the observable CBC merger rate, normalized to the latest limits from the LVK collaboration. The red and green dotted lines denote the most probable BNS and BLMBH merger rates that would be inferred, respectively, using the Bayes factors. Note the different range on the horizontal axes: with LIGO A+ reliable source classification becomes challenging beyond $\sim120$ Mpc due to reduced SNR, while for ET this occurs beyond $\sim1200$ Mpc. For this analysis, we adopt the SLy EoS as a conservative benchmark; stiffer EoSs enhance distinguishability.}
	\label{fig: modified merger rate}
\end{figure}

The above rate density can be normalized to LVK's estimate for the same, viz., $10-1700\rm\,Gpc^{-3}\,yr^{-1}$, at the current time  $t=13.8$ billion years or $z=0$. Now, as remarked earlier, this normalization is subject to there being no other contribution to the CBC rate in the $1-2.5\,M_{\odot}$ except BNS mergers. In reality there can be BLMBH signals, which are non-distinguishable from BNSs beyond a certain $D_L$. If we assume there are only two candidates for a low-mass CBC, viz.,  BNS mergers  \emph{\&}  BLMBH mergers, then for a fixed $\rho_{\rm opt}$, the Bayes factor $\mathcal{B}\equiv\mathcal{B}^{\rm BNS}_{\rm BLMBH} = \mathcal{B}_{\rm BNS}^{\rm BLMBH}$. Thus, we find the probabilities of these CBCs being BNS or BLMBH mergers as
\begin{equation}
	p({\rm BNS|data})= \frac{\mathcal{B}}{1+\mathcal{B}}\quad {\rm and} \quad p({\rm BLMBH|data})= \frac{1}{1+\mathcal{B}}\,.
	\label{eq:modifiedbns}
\end{equation}
For small values of $D_L$, the Bayes factor remains sufficiently large to favor one model over the other. As $D_L$ increases and $\mathcal{B}$ starts decreasing, eventually tending to 1, the model probabilities approach $1/2$, indicating indistinguishability. In Fig.\,\ref{fig: modified merger rate} we show the theoretical CBC rate (Eq.\,\ref{eq:BNSrate}, solid blue line) as a function of the luminosity distance. The rate has been normalized using LVK's current estimate of the total CBC rate\,\cite{KAGRA:2021duu}. Of the three different statistical models mentioned in\,\cite{KAGRA:2021duu}, here we have chosen the Multi-Source Model (MS), which predicts the rate to be within $130-1700\,\rm Gpc^{-3}yr^{-1}$, and for the normalization at $z=0$, we choose the mean of this model, given as $915\,\rm Gpc^{-3}yr^{-1}$.

Up to a source distance of approximately 400 (25) Mpc, the Bayesian evidence remains sufficiently high to reliably distinguish between BNS and BLMBH mergers with ET (LIGO A+), implying that the inferred BNS merger rate is consistent with the true rate. However, beyond 400 (25) Mpc, the Bayesian evidence gradually decreases, and the probability of misidentifying BLMBH mergers as BNS mergers (or vice versa) increases. Beyond $\sim1200$ (120) Mpc for ET (LIGO A+), the distinction becomes difficult, and the two types of sources appear nearly identical. As a result, the inferred BNS merger rate may be overestimated. As noted earlier, stiffer EoSs enhance the ability to distinguish BNS from BLMBH mergers, enabling reliable classification even at larger $D_L$ and resulting in tighter constraints.

\subsubsection{Exclusion Sensitivity for $f_{\rm BLMBH}$}
\label{sec:flmbh}
Unlike the previous section, where we estimated the rates, now we ask a different question: What is the fraction $f_{\rm BLMBH}$ of the CBC rate that can be compatible with being BLMBHs? To answer this question, we assume that we are given the detector specifications and the true model where all CBCs are in fact BNSs.

The number of CBCs expected to be detected ($N_{\rm CBC}$) is\,\mbox{\cite{Taylor:2011fs, Taylor:2012db, Finn:1992wt}},
\begin{align}
	N_{\rm CBC}&= T\times \int_0^{\infty} dz\, \frac{4\pi D_c^2(z)}{(1 + z) H(z)}\, C_\Theta\,R_{\text{CBC}}(z)\,,
	\label{eq: detectionrate}
\end{align}
where $T$ is the years of observation,  $R_{\rm CBC}$ is defined in Eq.\,\ref{eq:BNSrate}, $D_c(z)$ denotes the comoving radial distance as a function of redshift, and $H(z)$ is the Hubble expansion rate at a redshift $z$. The cosmological parameters are taken from the latest Planck measurements\,\cite{Planck:2018vyg}. The angular dependence of the SNR is encoded within $C_\Theta$ that varies within 0 and 4, calculated from the geometry of the sources and the antenna\,\mbox{\cite{Taylor:2011fs, Finn:1995ah}}. Apart from $R_{\rm CBC}$, the remainder of the expression in Eq.\,\ref{eq: detectionrate} denotes the exposure $\langle \rm VT \rangle$ for a specific detector. See Appendix \ref{sec:A2} for the computation of the detector exposures for LIGO A+ and ET.

\subsubsection*{Projected Exclusion Significance}
%\label{sec:A3}
We now perform a hypothesis test. In each independent redshift bin $i$,  the total number of observed BLMBH events is modeled as a Poisson variable with mean $\langle n_i \rangle = \mu S_i + B_i$, where $S_i$ is the expected number of BLMBH merger events, and $B_i$ is the expected number of background BNS events that are misclassified as BLMBH events, and $\mu$ is a hypothesis label with $\mu = 1$ representing the presence of BLMBH signals and $\mu = 0$ corresponding to the null (background-only) hypothesis. 

The total number of events in the $i^{\rm th}$ bin, from a Poisson process is given by
\begin{align}
	P(n_i)=\frac{( \mu S_i+B_i)^{n_i} e^{-(\mu S_i+B_i)}}{n_i!}\,,
	\label{eq: Poisson}
\end{align}
where $S_i$ \emph{\&} $B_i$ in our case can be calculated using Eq.\,\ref{eq: detectionrate}.
The log-likelihood ratio of the signal hypothesis ($\mu=1$) over background-only hypothesis ($\mu=0$) is given by
\begin{align}
	\nonumber
	q&=-2{\rm\,ln} \left(\frac{\mathcal{L}(S|_{\mu=1})}{\mathcal{L}(\hat{S}|_{\mu=0})}\right)\\
	&= -2 \sum_i\left(B_i\, {\rm ln}\frac{S_i+B_i}{B_i}-S_i\right)\,.
	\label{eq:likelihoodratio}
\end{align}

The median expected exclusion significance $ Z $, i.e. the ``Asimov'' significance for rejecting the signal-plus-background hypothesis in against the background-only hypothesis, is given by\,\mbox{\cite{ParticleDataGroup:2024cfk, Cowan:2010js}}
\begin{align}
	{Z}=\sqrt{q}= \sqrt{2\sum_i\left(B_i\, {\rm ln}\frac{B_i}{S_i+B_i}+S_i\right)}\,.
	\label{eq:significance}
\end{align}
A threshold of $ Z \geq 1.28 $ corresponds to a 90\% confidence level. In the left panel of Fig.\,\ref{fig:constraints_fLMBH}, we show how $Z$ depends on $f_{\rm BLMBH}$, when all events up to 450 Mpc are included at LIGO A+ and ET. Values of $f_{\rm BLMBH}$ larger than the shown threshold are expected to get ruled out.

\begin{figure}[t]
	\centering
	\includegraphics[width=0.444\textwidth]{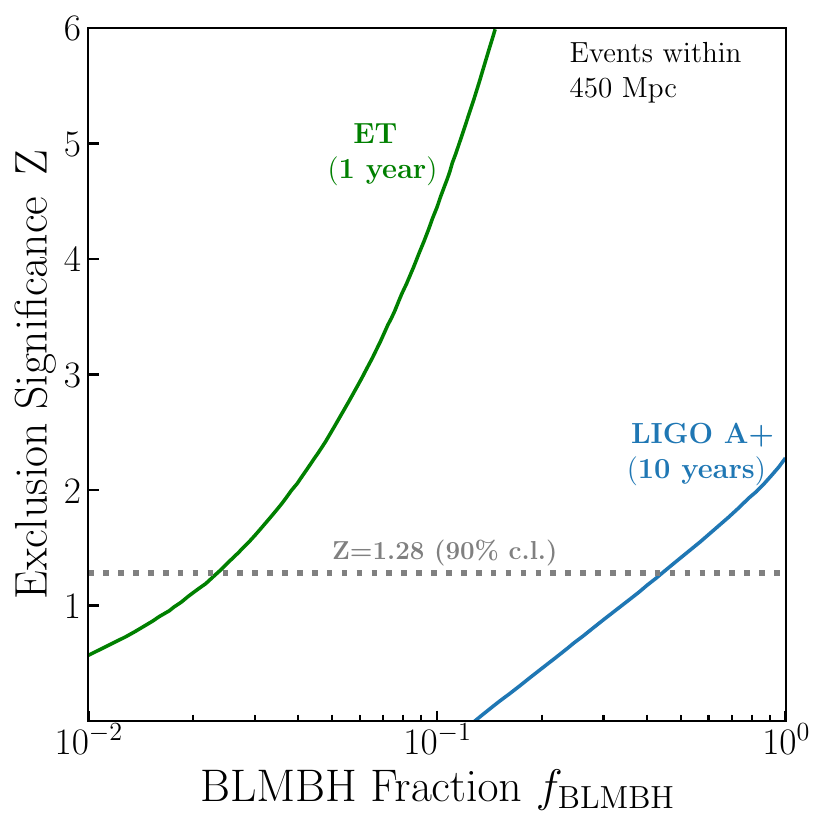}
	\hspace{0.5 cm}
	\includegraphics[width=0.47\textwidth]{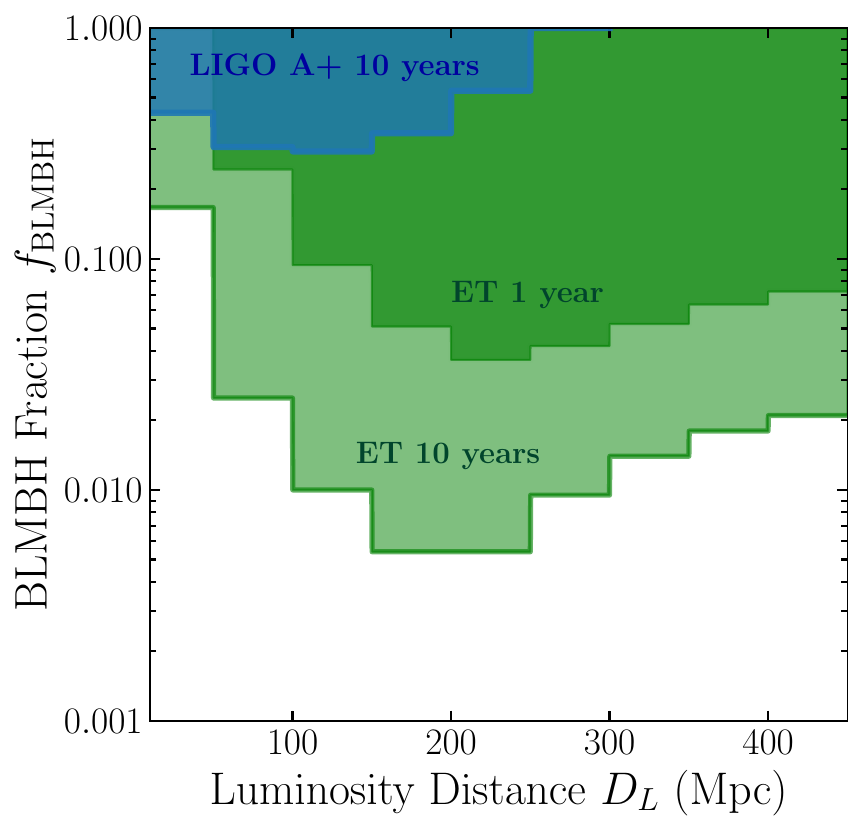}
	\caption{Left panel: Exclusion significance $Z$ as a function of BLMBH fraction $ f_{\rm BLMBH} $. For larger $ f_{\rm BLMBH}$, the number of BLMBH-induced signal events rises relative to the background, leading to a stronger rejection of the BLMBH+BNS hypothesis over the true BNS-only hypothesis. \change{Here we  include all events up to 450 Mpc.} Right panel: 90\,\% exclusion sensitivity to the fraction of BLMBH mergers in CBCs observable at  LIGO A+ and the ET, in nine luminosity distance bins, each of width 50\,Mpc. The constraints improve with increasing distance due to enhanced exposure (see Fig.\,\ref{fig_exposure}), but begin to weaken beyond $ \sim 250\,\mathrm{Mpc} $ (for ET) as the distinguishability starts to decrease, reducing the ability to reject background BNS events. \change{The $\rm FF$ and the $\mathcal{B}$ used in these two figure panels were computed using the soft EoS SLy to be conservative; stiffer EoSs will give more promising results.}}
	%\label{fig_significance}
	\label{fig:constraints_fLMBH}
\end{figure}

If a fraction $f_i$ of the CBCs are LMBHs by origin, then the expected signal (BLMBH) and background (BNS) events in the $i^{\rm th}$ redshift bin are given by Eq.\,\ref{eq: detectionrate} as
\begin{align}
	\nonumber
	&S_i = f_i\times R_{\text{CBC},i} \, \langle {\rm VT} \rangle_i \, \frac{\mathcal{B}_i}{1+\mathcal{B}_i}\,, \\
	&B_i =\left(1-f_i\right) \times R_{\text{CBC},i} \, \langle {\rm VT} \rangle_i \, \frac{1}{1+\mathcal{B}_i}\,,
	\label{eq:sigback}
\end{align}
where $\mathcal{B}=\mathcal{B}^{\rm LMBH}_{\rm BNS}=\mathcal{B}^{\rm BNS}_{\rm LMBH}$, assuming a fixed $\rho_{\rm opt}$. 
We can now compute the test statistic $Z$ in each bin and obtain the expected upper limit on $f_i$.

In the right panel of Fig.\,\ref{fig:constraints_fLMBH}, we present the 90\% exclusion sensitivity to the BLMBH merger fraction, as a function of $D_L$, for both LIGO A+ and the ET. The analysis is performed in 9 bins, each with a width of 50 Mpc.  The blue shaded region corresponds to 10 years of LIGO A+ observations, while the green shaded regions corresponding to ET, shown for 1 year and 10 years of observation time. For this analysis, we assume a fiducial CBC rate of $R_{\rm CBC} = 1000\,\mathrm{Gpc}^{-3}\,\mathrm{yr}^{-1}$. This rate appears as an overall normalization and scales the resulting constraints proportionally. As expected, the sensitivity drops at higher distances (owing to lack of distinguishability) and at very small distances (owing to small number of events). We have again adopted a conservative choice of the EoS on the softer side, specifically the SLy EoS. Choosing a stiffer EoS would lead to more pronounced matter effects in the waveform, thereby yielding comparatively stronger sensitivities.

\subsection{Exclusion Sensitivity for DM-Nucleon Interaction}
\label{sec:TBH}
The preceding discussions are agnostic to the origin of LMBHs. However, now we specifically consider the scenario in which LMBHs are formed via DM capture-induced collapse of BNS systems\,\mbox{\cite{Bhattacharya:2023stq, Dasgupta:2020mqg, McDermott:2011jp, Goldman:1989nd}}. Galactic DM with non-gravitational interactions with nucleons ($\sigma_{\chi n}$, the interaction strength) can be captured by NSs over their lifetimes\,\mbox{\cite{Gould:1987ir, Gould:1987ww, Bertone:2007ae, McDermott:2011jp, Kouvaris:2018wnh, Joglekar:2019vzy, Dasgupta:2020mqg, Dasgupta:2020dik, Bell:2020jou, Acevedo:2020gro, Bramante:2023djs, Bhattacharya:2023stq, Ray:2023auh}}. Once captured, heavy DM can accumulate in the NS core, forming a dense dark core that may become gravitationally unstable and collapse into a microscopic BH. This BH can eventually consume the entire NS, ejecting approximately $10^{-3}-10^{-4}\,M_\odot$\,\cite{East:2019dxt}, and forming an LMBH of mass comparable to the original NS\,\mbox{\cite{ Goldman:1989nd, McDermott:2011jp, Kouvaris:2018wnh, Garani:2018kkd, Dasgupta:2020mqg,  Dutta:2024vzw, Liu:2024qbe}}. The BLMBH merger rate becomes directly dependent on the underlying BNS merger rate, and one can obtain limits on the DM-nucleon interaction cross section $\sigma_{\chi n}$. 

Considering a constant $\sigma_{\chi n}$, corresponding to a heavy mediator of the DM-nucleon interactions, the capture rate is given as\,\mbox{\cite{McDermott:2011jp,Garani:2018kkd, Bhattacharya:2023stq}},
\begin{align}
	&C = 1.4\times 10^{20}\,{\rm s}^{-1}\,\Big(\tfrac{\rho_{\chi}}{0.4\, \rm{GeV\,cm^{-3}}}\Big) \Big(\tfrac{10^5\, \rm{GeV}}{m_{\chi}}\Big)\Big(\tfrac{\sigma_{\chi n}}{10^{-45}\,\rm{cm^2}}\Big) \nonumber\\
	&\phantom{C = }\times\left(1-\tfrac{1-e^{-A^2}}{A^2}\right)\,\left(\tfrac{v_{\rm{esc}}}{1.9\times10^{5}\rm km\,s^{-1}}\right)^2\left(\tfrac{220\,\rm km\,s^{-1}}{\bar{v}_{\rm gal}}\right)\,.
\end{align}
The factor involving $A^2={6\,m_{\chi}m_n}{v^2_{\rm{esc}}}/{\bar{v}^2_{\rm gal}}{(m_{\chi}-m_n)^2}$ accounts for inadequate momentum transfers at larger $m_\chi$, given NS escape speed $v_{\rm esc}$ and typical DM density $\rho_{\chi}$ and speeds ${\bar{v}_{\rm gal}}$ in the galaxy. For a typical $1.35\,M_{\odot}$ NS, with 10 km radius, with a lifetime of 1 Gyr, and with a core temperature ($T_{\rm core}$) of $2.1\times 10^6\,\rm K$, the accumulated DM mass can reach up to $\sim 10^{-15}\,M_{\odot}$. If the captured DM particles are sufficiently massive, they can drift toward the center of the star and thermalize, forming a dense core within a thermal radius $r_{\rm th}$. For example, DM with $m_{\chi} = 10^5$\,GeV can form such a core within a radius of approximately 5\,cm\,\cite{Bhattacharya:2023stq}. This overdense core may eventually undergo gravitational collapse, either due to self-gravitation or due to overcoming the Fermi degeneracy pressure, depending on whether the DM particle is bosonic or fermionic\,\mbox{\cite{McDermott:2011jp, Dasgupta:2020mqg}}. The resulting collapse can lead to the formation of a small seed BH at the center of the NS. Once formed, the seed BH at the core of the NS can eventually consume the entire star, leading to the formation of a black hole with mass comparable to that of the host star\,\cite{East:2019dxt}. Such LMBHs, originating from DM capture-induced collapse of NSs, are referred to as transmuted black holes\,(TBHs).

\begin{figure}[t]
	\centering
	\includegraphics[width=0.325\textwidth]{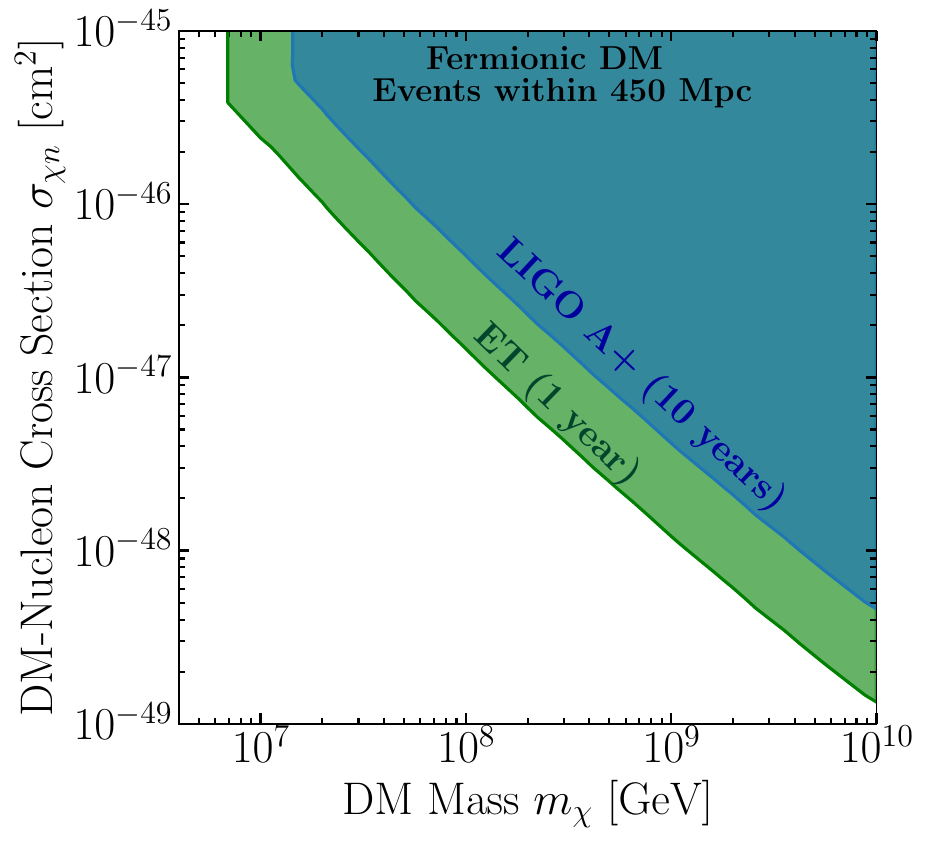}
	\includegraphics[width=0.325\textwidth]{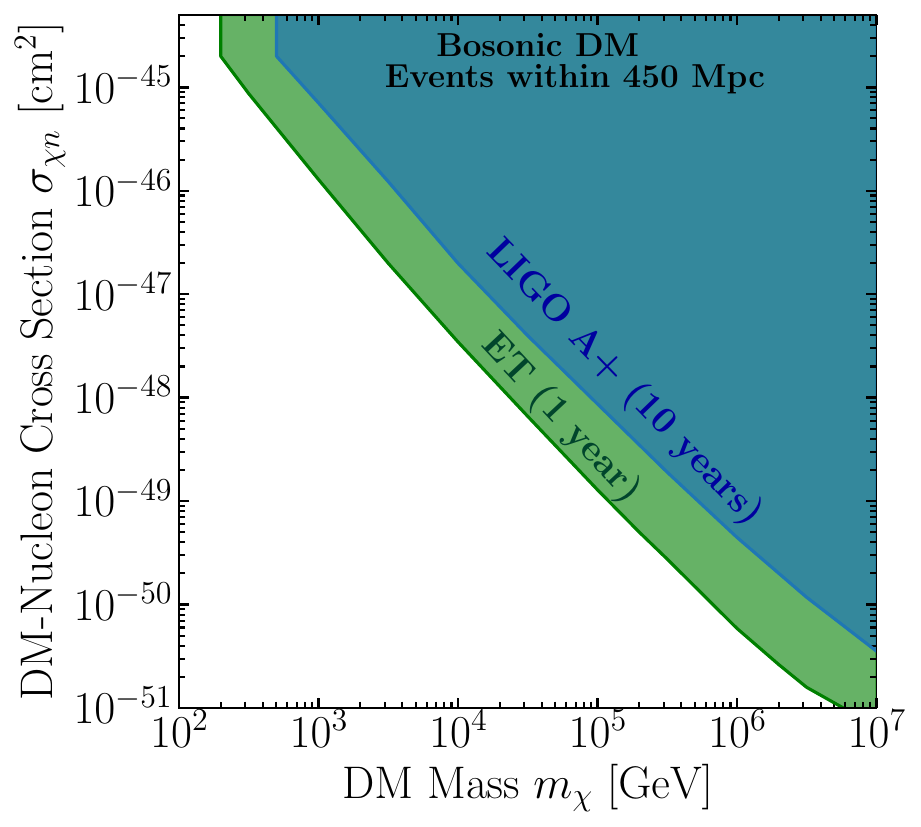}
	\includegraphics[width=0.325\textwidth]{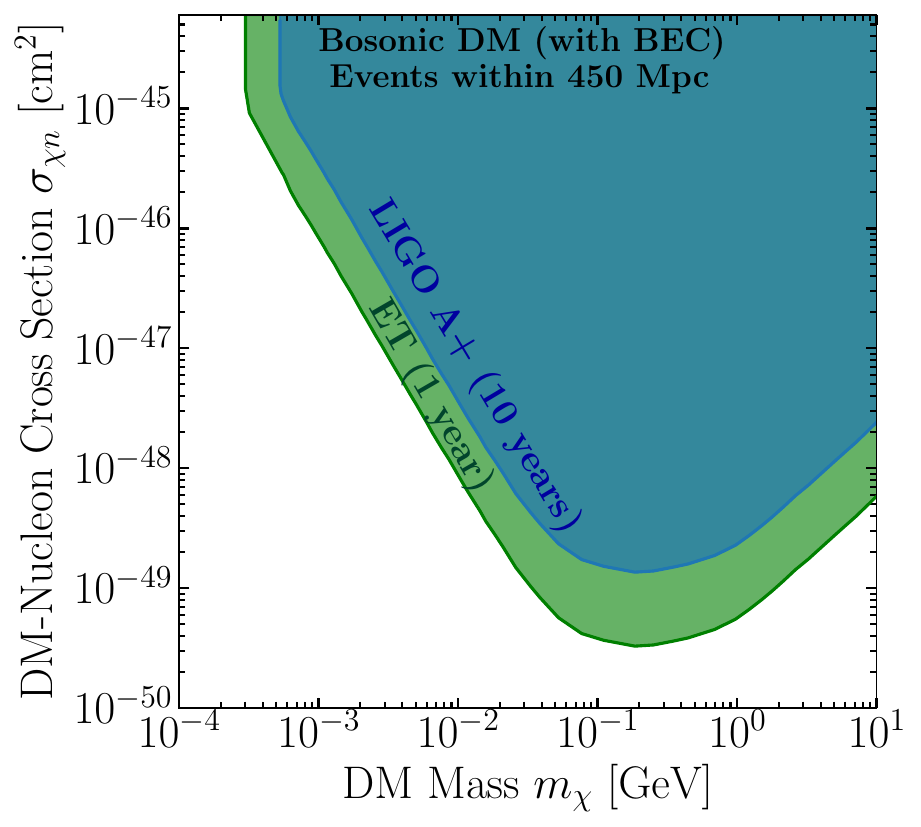}
	
	\caption{Exclusion sensitivity for DM-nucleon cross section $\sigma_{\chi n}$ as a function of DM mass $m_\chi$. This analysis holds for non-annihilating DM (of three types: fermionic, bosonic, and bosonic with BEC formation), and includes all events within $D_L \leq 450\,\mathrm{Mpc}$. We adopt a fiducial merger rate of $R_{\rm CBC} = 1000\,\mathrm{Gpc}^{-3}\,\mathrm{yr}^{-1}$, which acts as a normalization factor for the limits. The distinguishability between BLMBH mergers and BNS mergers is computed assuming a soft nuclear EoS for the BNS mergers (SLy); adopting a stiffer EoS would yield stronger constraints. Competing limits can be found in\,\cite{Bhattacharya:2023stq} where the limit was calculated assuming a zero background (no misclassified events) and a uniform prior on the CBC rate within $10-1700\rm Gpc^{-3}\,yr^{-1}$.
	}
	\label{fig:constraints_DM}
\end{figure}

The merger rate of these TBH-TBH binaries\,\cite{Dasgupta:2020mqg} is given as:
\begin{align}
	R_{\rm TBH}= \int dr\frac{df}{dr}\int_{t_*}^{t_0}dt_f\frac{dR_{\rm CBC}}{dt_f}\times \Theta \left[t_0-t_f-\tau_{\rm trans}\left[m_{\chi},\sigma_{\chi n}, \rho_{\rm ext}(r,t_0)\right]\right]\,,
	\label{eq:RTBH}
\end{align}
where ${df}/{dr}$ is the radial distribution of the progenitor BNSs in a galaxy. The TBH merger rate intrinsically depends on the BNS merger rate, as only a fraction of originally BNS systems can evolve into TBH binaries. This fraction is captured by the step function $\Theta\left[t_0 - t_f - \tau_{\rm trans}(m_{\chi}, \sigma_{\chi n}, \rho_{\rm ext}(r, t_0))\right]$, which encodes the condition that the transmutation timescale $\tau_{\rm trans}=\tau_{\rm collapse}+\tau_{\rm swallow}$, which is a function of the dark matter mass $m_{\chi}$, nucleon scattering cross section $\sigma_{\chi n}$, and ambient dark matter density $\rho_{\rm ext}$, must be shorter than the time available since binary formation.  Table 1 of Ref.\,\cite{Bhattacharya:2024cpm} lists the timescales relevant to this scenario. The timescale $\tau_{\rm collapse}$ -- associated with the seed BH formation from collapse of the accreted DM and depending on the nature of the DM particles: fermionic DM, bosonic DM without Bose-Einstein condensate (BEC) formation, or with BEC formation -- is shown in Eq.\,\ref{eq:collapsetime}.
\begin{align}
	\begin{aligned}
		\tau_{\rm collapse}^{\rm fermion} &=
		1.9\times 10^{10}\,{\rm years}
		\left(\tfrac{10^8\,{\rm GeV}}{m_\chi}\right)
		\left(\tfrac{0.4\, {\rm GeV\, cm^{-3}}}{\rho_\chi}\right)
		\left(\tfrac{10^{-45}\,{\rm cm^2}}{\sigma_{\chi n}}\right), \\[6pt]
		\tau_{\rm collapse}^{\rm boson} &=
		4.8\times 10^8 \,{\rm years}
		\left(\tfrac{T_{\rm core}}{2.1\times 10^6\, {\rm K}}\right)^{3/2}
		\left(\tfrac{10^5\,{\rm GeV}}{m_\chi}\right)^{3/2}
		\left(\tfrac{0.4 \,{\rm GeV\, cm^{-3}}}{\rho_\chi}\right)
		\left(\tfrac{10^{-45}\,{\rm cm^2}}{\sigma_{\chi n}}\right), \\[6pt]
		\tau_{\rm collapse}^{\rm BEC} &=
		1.1\times 10^9 \,{\rm years}
		\left(\tfrac{10^{-2}\,{\rm GeV}}{m_\chi}\right)^{2}
		\left(\tfrac{0.4 \,{\rm GeV\, cm^{-3}}}{\rho_\chi}\right)
		\left(\tfrac{10^{-45}\,{\rm cm^2}}{\sigma_{\chi n}}\right).
	\end{aligned}
\label{eq:collapsetime}
\end{align}
In addition, the time taken for the seed BH to consume its host star is $\tau_{\rm swallow}\approx 3\times10^{4}\,{\rm yr}\,(M_{\rm seed\,BH}/M_\odot)$; this depends on the DM mass through $M_{\rm seed\,BH}$ and can exceed the $\tau_{\rm collapse}$ for heavier DM. See Ref.\,\cite{Bhattacharya:2023stq} for a detailed computation of these timescales, including possible uncertainties.

TBH-TBH mergers could be observable in GW detectors as BLMBH mergers. The absence of such detections thus places stringent constraints on the DM-nucleon scattering cross section $\sigma_{\chi n}$\,\mbox{\cite{Bhattacharya:2023stq, Singh:2022wvw}}. However, the constraints in Ref.\,\cite{Bhattacharya:2023stq} were obtained under the assumption that no BLMBHs would be actually detected. In the language of this paper, this is akin to assuming that no BNS mergers would be misclassified as BLMBH mergers. 
The method presented in this paper can be used to assess the exclusion sensitivity to the DM parameter space \emph{without} the above assumption. Interpreting the LMBHs as TBHs (Eq.\,\ref{eq:RTBH}), we translate our limits on $f_{\rm BLMBH}$ into bounds on the DM mass and its scattering cross section with nucleons, as shown in Fig.\,\ref{fig:constraints_DM}. For this purpose, we replace $f \times R_{\rm CBC}$ in Eq.\,\ref{eq:sigback} with the TBH-TBH merger rate $R_{\rm TBH}$ (Eq.\,\ref{eq:RTBH}), and determine the region in the $(m_{\chi}-\sigma_{\chi n})$ plane where the statistical significance $Z$ is $\geq 1.28$. 

In Fig.\,\ref{fig:constraints_DM} we show projected exclusion on the DM parameter space.  For this analysis, we integrated up to a maximum luminosity distance of $D_L = 450\,\mathrm{Mpc}$ to compute the total number of signal and background events. The resulting bounds thus reflect the contribution from all the nine bins up to 450 Mpc. The exclusion sensitivity of Fig.\,\,\ref{fig:constraints_DM}, for 10 years observation with LIGO A+, is slightly stronger than the projected upper limits shown in Fig.\,1 of Ref.\,\cite{Bhattacharya:2023stq}. This is simply because the exposure $\langle\rm VT\rangle$ in this case is approximately twice the exposure used in the forecasted limit in Ref.\,\cite{Bhattacharya:2023stq}. ET can achieve comparable (or stronger) exclusions with 1 year of data. More broadly though, for the observation horizon considered, misclassification does not lead to a drastic degradation of the previously projected limits.

\section{Discussions \emph{\&} Conclusion}
\label{sec:discuss}
In this study, we have systematically investigated the potential degeneracy between BNS mergers and BLMBH mergers. As a first step, we considered systems with identical component masses and spin configurations for both BNS and BLMBH binaries, focusing on the $1.35\,M_{\odot}$-$1.35\,M_{\odot}$ case. While the BLMBH waveform can be obtained directly from standard general relativistic BBH computations, generating accurate BNS waveforms is considerably more challenging due to the involvement of complex microphysical processes, most notably the dependence on the NS EoS. To capture this effect, we extracted BNS waveforms from the publicly available CoRe database, carefully selecting eight different EoSs ranging from the stiffest to the softest cases, and analyzed their impact on the late inspiral and postmerger phase waveforms.

After obtaining the two sets of waveforms, we analyzed the mismatch between them. We found that the dominant mismatch arises in the high-frequency regime, as matter effects become significant in higher frequency signals from the late inspiral and postmerger phases. This behavior is consistent with the expectation that the microphysical properties of NS matter, governed by the EoS, manifest in the GW signal predominantly in the strong-field regime. We divide the waveforms into two distinct phases: the late inspiral and the postmerger. Through this separation, we demonstrate that the fit between the BNS and LMBH waveforms deteriorates significantly in the postmerger phase compared to the inspiral. The EoS-dependent effects lead to characteristic differences in the emitted GW signal. We compute the Bayesian evidence in favor of BNS models over BLMBH models, across a variety of NS EoSs. Our analysis highlights the critical role of detector's sensitivity across low and high frequencies, which influences the ability to distinguish between LMBH and BNS mergers and which part of the signal gives more information. While natively the strongest differences lie in the postmerger phase (and detectors like NEMO can leverage it), the proposed detectors CE and ET are more sensitive at lower frequencies and the inspiral signal provides greater sensitivity therein. However, this conclusion may be altered by degeneracies with other parameters such as spins and eccentricities. The variation in Bayesian evidence with distance has direct implications for key astrophysical measurements, most notably, the estimation of the BNS merger rate. Misidentification of LMBH mergers as BNS events at large distances can lead to an overestimation of the true BNS rate, thereby impacting theoretical modeling, and population studies.

We then computed expected exclusions on the maximal contribution of BLMBH mergers to the observed CBC rate. We perform a thorough statistical analysis incorporating BLMBH merger signals and background contributions arising from the misclassification of BNS mergers, particularly in regimes where waveform degeneracies occur due to limited detector sensitivity.
We further consider the scenario in which these BLMBHs originate from DM capture-induced transmutation of BNSs. In this context, the limits on $ f_{\rm BLMBH} $ can be translated into constraints on the DM parameter space, specifically the DM mass $ m_{\chi} $ and its scattering cross section with nucleons\,$\sigma_{\chi n}$. These exclusions are similarly based on rejecting the BLMBH+BNS hypothesis in favor of the BNS-only hypothesis, i.e. the absence of a true BLMBH component in the low-mass CBC sample. 
We intentionally exclude the case of mixed BLMBH-NS binaries in this study, as one of our motivations was BLMBHs formed via dark matter capture-induced transmutation. Given that the binary neutron stars are typically close enough to reside in the same ambient DM environment, it is expected that both components undergo transmutation, resulting in symmetric LMBH-LMBH systems.

To summarize, we have studied the possibility to distinguish BNSs from BLMBHs, and constrain exotic LMBH formation channels, including DM-induced scenarios such as TBH formation. Unlike prior studies relying on inspiral-only waveforms, our analysis spans both inspiral and postmerger regimes and quantifies the distinguishability using Bayesian evidence. This has implications for dark matter constraints and BNS rate estimates, and highlights the need for improved high-frequency modeling\,\cite{Aggarwal:2025noe} for probing NS EoSs and a variety of exotic physics scenarios\,\cite{Harada:2023eyg}.

%\section*{Note Added}
%After our work was completed, we became aware of the preprint \href{https://arxiv.org/abs/2507.07895}{arXiv:2507.07895}\,\cite{Khadkikar:2025awf}. The above preprint explores tidal effects in the inspiral phase of GW190425-like systems as a means to distinguish NSs from BHs. In contrast, our work presents the first systematic study of the postmerger regime, incorporating a broad set of EoSs and leveraging the information from the second peak beyond the inspiral. While the above preprint advocates for and uses a Savage-Dickey ratio to quantify distinguishability, appropriate in the case of nested hypotheses, we use the more generally applicable Bayes factor. More broadly, while these two studies have a similar aim, they differ significantly in focus, scope, and detail. Our work has been presented previously, e.g., at \href{https://indico.global/event/8004/contributions/72171/}{PPC 2024} in IIT Hyderabad, at \href{https://www.icts.res.in/program/gwbsm2024/talks}{GW BSM 2025} in ICTS-TIFR Bengaluru, and at \href{https://indico.math.cnrs.fr/event/12305/contributions/13054/}{NDM 2025} in IHP Paris. 

\acknowledgments
We thank Parameswaran Ajith, Aryaman Bhutani, Prolay Krishna Chanda,  Ranjan Laha, David Radice, Anupam Ray, and Aditya Vijaykumar for helpful discussions and suggestions while this manuscript was being prepared. BD thanks Reed Essick, Maya Fishbach, and Philippe Landry at CITA for hospitality during the summer of 2023 and for helpful conversations that motivated this study. This work is supported by the Dept. of Atomic Energy (Govt. of India) research project RTI 4002, and by the Dept. of Science and Technology (Govt. of India) through a Swarnajayanti Fellowship to BD. BD and SB acknowledge the support of the Institut Henri Poincaré (UAR 839 CNRS-Sorbonne Université), and LabEx CARMIN (ANR-10-LABX-59-01) during the ``NDM 2025'' workshop. SB thanks the organizers of N3AS summer school on Multi-messenger Astrophysics for hospitality.
%\newpage
\appendix
\section{Appendices}
In this appendix we briefly review definitions, data sources, and a number of auxiliary quantities needed to reproduce the analysis and results shown in the main text. 

\subsection{BLMBH Waveforms from PyCBC}
\label{sec:LMBH waveform}
\begin{figure}[h]
	\centering
	\includegraphics[width=0.8\textwidth]{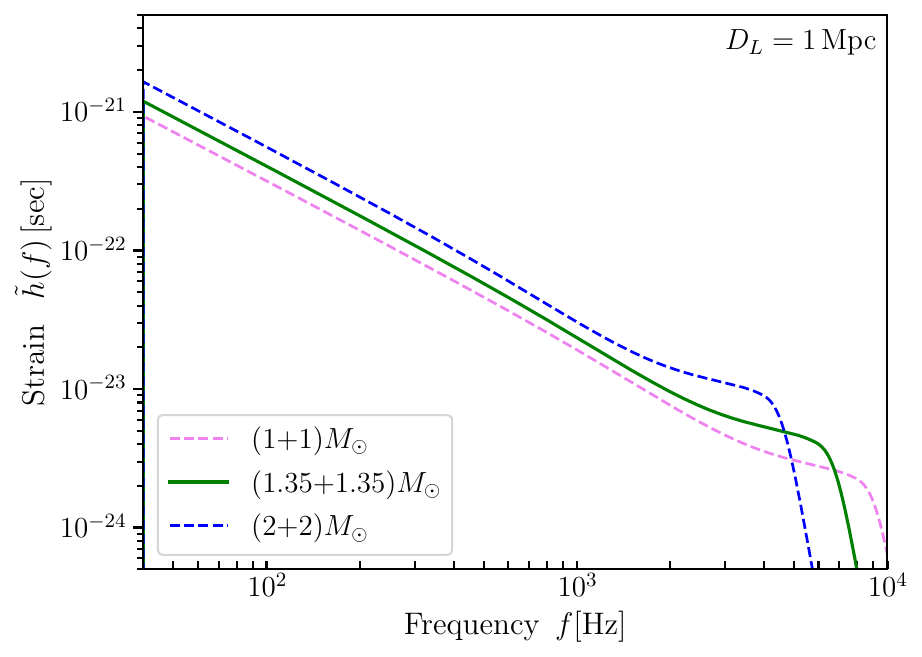}
	\caption{BLMBH merger waveform, strain $\tilde{h}(f)$ vs. frequency~($f$), generated by IMRPhenomD\,\mbox{\cite{Husa:2015iqa, Khan:2015jqa}} with total masses $2M_{\odot}$, 2.7$M_{\odot}$, and 4$M_{\odot}$ respectively, with vanishing spin components, and with $D_L=$\,1 Mpc. The strain falls as $\propto\, f^{-7/6}$ during the inspiral and scales as $\propto\,M_c^{5/6}$, as shown in Eq.\,\ref{eq_strain}.}
	\label{fig_BBH}
\end{figure}

\change{We generate the LMBH-LMBH merger waveform} by using the publicly available software PyCBC\,\cite{alex_nitz_2024_10473621}. We employ its time domain and frequency domain module named IMRPhenomD\,\mbox{\cite{Husa:2015iqa, Khan:2015jqa}}, which is a phenomenological model of GW signals from the inspiral-merger-ringdown of non-precessing (aligned spin) black hole binaries. These waveforms \change{cover the entire inspiral, merger and ringdown regime, where the strong gravity effects are calibrated with numerical relativity (NR) simulations}, unlike the waveforms generated using perturbative expansion in $v/c$, where $v$ is the orbital velocity and $c$ is the speed of light. Fig.\,\ref{fig_BBH} gives the frequency domain waveforms of LMBH mergers for different component masses.

For all our BLMBH waveforms we consider two non-spinning LMBHs with equal masses, $m_1=m_2\in1-2.5\,M_{\odot}$, for which the chirp mass is \mbox{$M_c = \mu^{3/5}\,M^{2/5}$}, where $\mu\equiv{m_1 m_2}/(m_1+m_2)$ is the reduced mass of the system and $M\equiv\left(m_1+m_2\right)$ is the total mass of the system. While the IMRPhenomD waveforms we use are more detailed, below we describe the basic features of BLMBH waveforms.

The `plus' (+) and `cross'\,($\times$) polarized strain amplitudes for the dominant ($l=2$, $m=2$) mode are given by\,\cite{Maggiore:2007ulw}
\begin{align}
	\nonumber
	\label{eq_strain}
	&\tilde{h}_+(f) = \left(\frac{5}{24}\right)^{1/2}\frac{1}{\pi^{2/3}}\frac{1}{D_L} \left(GM_c\right)^{5/6}f^{-7/6}e^{i\Psi_+(f)} \left(\frac{1+\cos^2{\iota}}{2}\right),\\
	&\tilde{h}_{\times}(f) = \left(\frac{5}{24}\right)^{1/2}\frac{1}{\pi^{2/3}}\frac{1}{D_L} \left(GM_c\right)^{5/6}f^{-7/6}e^{i\Psi_{\times}(f)} \cos{\iota},
\end{align}
where $G$ is the gravitational constant, $D_{L}$ is the luminosity distance, $M_c$ is the detector-frame chirp mass given by $(1+z)M_c|_{\rm source\, rest\, frame}$. The redshift $z$ can be assumed to be 0 for current observations so far. The inclination of the perpendicular to the orbital plane with respect to the line of sight is given as ${\iota}$. If $\iota=\pi/2$, the dominant strain amplitude is reduced by a factor of 1/2 and $h_{\times}$ vanishes, resulting in a linearly polarized wave; this orientation is termed as `edge-on'. For $\iota= 0\, \emph{\&} \, \pi$, the system's orbital angular momentum is aligned with the line of sight and this is termed as a `face-on' orientation, resulting in a circularly polarized wave. For a face-on symmetric $\left(m_1=m_2\right)$ LMBH binary at $D_L=$\,1 Mpc (with vanishing spins), the strain amplitude will be
\begin{align}
	&\tilde{h}(f)\approx 4.2\times 10^{-22}\, {\rm sec} \left(\frac{\rm Mpc}{D_L}\right)\left(\frac{M_c}{1.17 M_{\odot}}\right)^{5/6}\left(\frac{f}{100\, \rm Hz}\right)^{-7/6}\,,
\end{align}
with the typical $f^{-7/6}$ scaling with frequency that is expected from the inspiral of two point masses.
 
This \change{analytical} waveform is valid up to a maximum frequency, termed as the frequency of innermost stable circular orbit (ISCO), given by 
\begin{align}
	f_{\rm ISCO} = \frac{1}{6\sqrt6 \left(2\pi\right)}\frac{c^3}{GM}\approx 2.2\, {\rm kHz}\,\left(\frac{M_{\odot}}{M}\right)\,.
\end{align}
Beyond the ISCO, the inspiral phase ends and the binary undergoes a rapid plunge, marking the onset of the merger phase, which is followed by a damped sinusoidal ringdown\,\mbox{\cite{Hughes:2019zmt, Berti:2009kk}} as the remnant approaches equilibrium.

\subsection{BNS Waveforms from CoRe Database \emph{\&} Choice of EoSs}
\label{sec: BNS waveform}
\begin{figure}[t]
	\centering
	\includegraphics[width=0.8\textwidth]{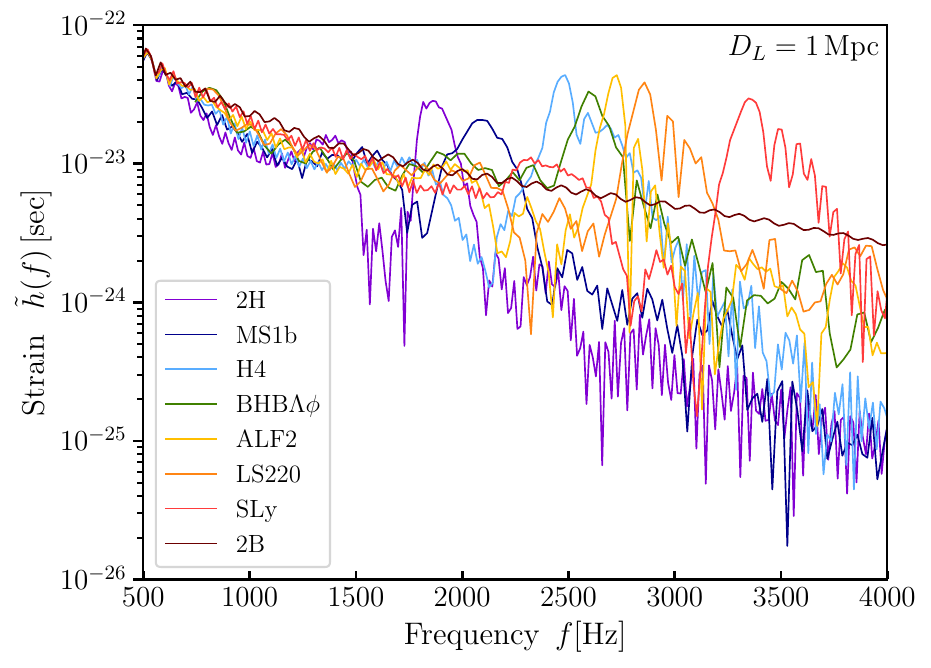}
	\caption{Frequency domain waveforms of $1.35\,M_{\odot}-1.35\,M_{\odot}$ BNS mergers, with non-spinning components, for the EoSs described in Table~\ref{tab:fmerge}.}
	\label{fig_FD_BNS}
\end{figure}

The Computational Relativity (CoRe) database\,\cite{CoreWebsite} provides BNS merger waveforms obtained \change{by solving the Einstein equations numerically.}  We have focused on symmetric $m_1=m_2=1.35M_{\odot}$ BNS systems, mimicking the observed Galactic binary neutron star mass distribution that is approximately a Gaussian between $1.08-1.57\,M_{\odot}$, with a mean at 1.35$M_{\odot}$\,\cite{Ozel:2016oaf}. When components of the BNS mergers are far apart, their dynamics can be approximated using point-particle methods similar to BBHs. However, near the merger and postmerger, non-linear hydrodynamics, neutrino transport, magnetic fields, and dense nuclear microphysics become essential to accurately model tidal interactions and matter ejection, necessitating fully numerical relativistic simulations\,\mbox{\cite{Dudi:2021wcf, Samajdar:2018dcx, Samajdar:2019ulq, Dietrich:2017aum, Dietrich:2020eud}}.

The first release of CoRe\,\cite{Dietrich:2018phi} comprised 367 waveforms derived from numerical simulations based on general relativity, employing 164 distinct setups with varying parameters such as total binary mass, mass ratio, initial separation, eccentricity, and stellar spins. The second release\,\cite{Gonzalez:2022mgo} included 254 distinct BNS configurations, resulting in a total of 590 numerical-relativity simulations performed at various grid resolutions. These simulations encompass a wide range of possibilities, with 18 different EoSs and a wide range of parameters. These simulations are executed with two independent mesh-based methods: BAM\,\mbox{\cite{Bruegmann:2006ulg,Thierfelder:2011yi}} and THC\,\cite{Radice:2012cu}.  BAM is better adapted to treat the NS surface, multiple orbits in the inspiral, followed by the merger and postmerger. Whereas THC is preferentially used to address the microphysics of the NS, as it contains different neutrino transport schemes, different mixing and dissipation mechanisms during the merger. For this work, we chose 6 EoSs from BAM simulations and 2 from THC simulations. 

The extraction of BNS waveforms from this database is discussed in detail (with the file structure and nomenclature) in the next section. This data is in time-domain, and we perform a Fourier transform to obtain the waveform in frequency domain. A sample of BNS waveforms from CoRe is shown in Fig.\,\ref{fig_FD_BNS}. They roughly agree with the BLMBH waveforms at low frequencies, but exhibit qualitative differences at higher frequencies due to the matter effects. \change{A characteristic secondary peak associated with the fundamental oscillation mode of the remnant is noticed in the postmerger spectrum of the frequency domain.\,\cite{Bauswein:2015vxa}} Depending on the EoS \change{and the maximum mass of the NS allowed by that EoS}, the BNS merger remnant may result in a hypermassive NS that subsequently collapses to a BH, or undergo a prompt collapse~\cite{Hotokezaka:2011dh, Bauswein:2013jpa}. The frequency of this second peak (and potentially more such peaks) depends on the details postmerger. However, note the second peak seen for all the EoSs, at $f\lesssim 4$\,kHz, except for 2B which is in tension with NS mass-radius data\,\cite{Gonzalez:2022mgo}.

The model parameters of this simulation are the masses of the component NSs, $m_1$ and $m_2$, and the effective spin parameter, 
\begin{align}
	\chi_{\rm eff} = \frac{m_1\chi_1 +m_2\chi_2}{M}\,,
\end{align}
where $\chi_i$ for $i^{\rm th}$ NS represents the spin component aligned with the angular momentum vector, and $M=m_1+m_2$ is the total binary mass. The tidal coupling constant and the reduced tidal parameters\,\mbox{\cite{Damour:2009wj, Favata:2013rwa}} are defined as
\begin{align}
	\nonumber
	&\kappa_2^T=3\nu\left[\left(\frac{m_1}{M}\right)^3\Lambda_1+\left(\frac{m_2}{M}\right)^3\Lambda_2\right],\\
	& \tilde{\Lambda}=  \frac{16}{13}\left[\frac{\left(m_1+12m_2\right)m_1^4\Lambda_1}{M^5}+\frac{\left(m_2+12m_1\right)m_2^4\Lambda_2}{M^5}\right]\,,
	\label{eq:eos}
\end{align}
where $\nu = {m_1m_2}/{M^2}$ is called the symmetric mass ratio. For $m_1=m_2=m$ it is 1/4. $\Lambda_i = \frac{2}{3}\kappa_{2,i}C_i^{-5}$ is the tidal polarizability parameter for the $i^{\rm th}$ NS, where $\kappa_{2,i}$ \emph{\&} $C_i$ are the gravito-electric Love number and compactness of the $i^{\rm th}$ NS respectively. With these inputs, the radiated gravitational wave decomposed in $\left(l,m\right)$ multipoles is
\begin{align}
	h_+-ih_{\times}=D_{L}^{-1}\sum_{l=2}^{\infty}\sum_{m=-l}^{l}h_{lm}(t)_{\,-2}Y_{lm}(\iota,\phi)\,,
\end{align}
where $_{-2}Y_{lm}$ are the $s = -2$ spin-weighted spherical harmonics\,\cite{Maggiore:2007ulw}.  The inclination and orbital phase, $\iota$ and $\phi$, respectively, give the inclination of the perpendicular to the orbital plane with respect to the line of sight, and the orbital phase of the binary at a particular time. We will only be interested in the $(2,2)$ mode, as for BLMBHs.

CoRe simulations use 18 different EoS models, including finite temperature EoS in the first release. BAM simulations are performed with analytical EoS in the form
\begin{align}
	P\left(\rho,\epsilon\right)= P_{\rm pwp}\left(\rho\right)+\left(\gamma_{\rm th}-1\right)\rho\left(\epsilon-\epsilon_{\rm pwp}\right)\,,
\end{align}
where $P_{\rm pwp}(\rho)$ is a given piecewise polytropic EoS model\,\cite{Read:2008iy}. The specific parameters employed for the piecewise polytropic EoS are available on the CoRe website\,\cite{CoreWebsite} and they mimic the zero temperature EoS models described in\,\cite{Read:2008iy}. Fig.\,1 of Ref.\,\cite{Gonzalez:2022mgo} shows the $m_{\rm NS} - R_{\rm NS}$ diagram and the $\Lambda- m_{\rm NS}$ diagram of these EoS models, which gives an overview of the current parameter space of EoS models. For a given mass, $M=1.4M_{\odot}$, the EoS 2H furnishes the largest radius $R^{\rm TOV}$ $\approx$\,15.21 km, and EoS 2B the smallest $\approx$\,9.75 km\,\mbox{\cite{ Gonzalez:2022mgo,Lattimer:2012nd}}. Evidently 2H is the stiffest (high pressure for a given density) and 2B is the softest (lower pressure for a given density) EoS in this database. 

\subsubsection*{Data Extraction from CoRe Database}
\label{sec:A1}
The CoRe\,\cite{CoreWebsite} database website provides a section named `GW-DB' where one can find the instructions about accessing the datasets required. One can also find information on the EoSs and a description of the data structure. The data for each waveform is stored in an HDF5\,\cite{HDF5Website} file, for example named as BAMXXXX.h5. Dismantling the file with h5py\,\cite{h5pyWebsite}, one can find the data structure of each waveform. The file is written in three groups:
\begin{enumerate}
	\item Energy --- contains the binding energy information of the binary in relation with reduced angular momentum\,\cite{Damour:2011fu}.
	\item $\rm rh_{22}$ --- contains the $l=m=2$ multipole of the metric waveform extracted at some coordinate radius r.
	\item  ${\rm r}\psi_4|_{lm}$ --- contains the Weyl curvature\,\cite{vanHolten:2022zsw} waveform up to $l=m=4$ multipole.
\end{enumerate}
As we will generate the BNS strain vs. frequency waveform for different EoSs, we are only concerned with the second group ($\rm rh_{22}$).

\begin{figure}[h]
	\centering
	\begin{tikzpicture}[every node/.style={align=center}, scale=0.9, transform shape]
		
		% Main BAM node
		\node[draw, text width=4cm] (bam) at (0, 0) {\textbf{BAM:0098}};
		
		% Category titles
		\node[draw, text width=5cm] (energy) at (-6, -3) {\textbf{Energy}};
		\node[draw, text width=5cm] (strain) at (0, -3) {\textbf{Strain ($rh_{22}$)}};
		\node[draw, text width=5cm] (weyl) at (6, -3) {\textbf{Weyl multipole} (${\rm r}\psi_4|_{lm}$)};
		
		% Arrows from BAM to category titles
		\draw[->] (bam) -- (energy);
		\draw[->] (bam) -- (strain);
		\draw[->] (bam) -- (weyl);
		
		% Entry group boxes (centered text, evenly spaced vertically)
		\node[draw, text width=5cm] (e_list) at (-6, -5.5) {
			\textnormal{1. EJ\_r00450.txt}\\[2pt]
			\textnormal{2. EJ\_r00600.txt}\\[2pt]
			\textnormal{3. EJ\_r00800.txt}\\[2pt]
			\textnormal{4. EJ\_r01000.txt}
		};
		
		\node[draw, text width=5cm] (s_list) at (0, -5.5) {
			\textnormal{1. Rh\_l2\_m2\_r00450.txt}\\[2pt]
			\textnormal{2. Rh\_l2\_m2\_r00600.txt}\\[2pt]
			\textnormal{3. Rh\_l2\_m2\_r00800.txt}\\[2pt]
			\textnormal{4. Rh\_l2\_m2\_r01000.txt}
		};
		
		\node[draw, text width=5cm] (w_list) at (6, -5.5) {
			\textnormal{1. Rpsi4\_lx\_my\_r00450.txt}\\[2pt]
			\textnormal{2. Rpsi4\_lx\_my\_r00600.txt}\\[2pt]
			\textnormal{3. Rpsi4\_lx\_my\_r00800.txt}\\[2pt]
			\textnormal{4. Rpsi4\_lx\_my\_r01000.txt}
		};
		
		% Straight vertical connectors from categories to their lists
		\draw[-] (energy.south) -- (e_list.north);
		\draw[-] (strain.south) -- (s_list.north);
		\draw[-] (weyl.south) -- (w_list.north);
		
	\end{tikzpicture}
	\caption{Structure of \texttt{BAM0098.h5} file. In the Weyl curvature multipole section, filenames are written as \texttt{Rpsi4\_lx\_my\_r00450.txt}, where \texttt{x} takes values from 2–4 and \texttt{y} from 0–4 depending on the $l$ values. \texttt{R} denotes the extraction radius where these quantities are calculated, with values 450$M_{\odot}$, 600$M_{\odot}$, 800$M_{\odot}$, and 1000$M_{\odot}$.}
	\label{BAM:0098}
\end{figure}
\begin{figure}[t]
	\centering
	\includegraphics[width=0.47\textwidth]{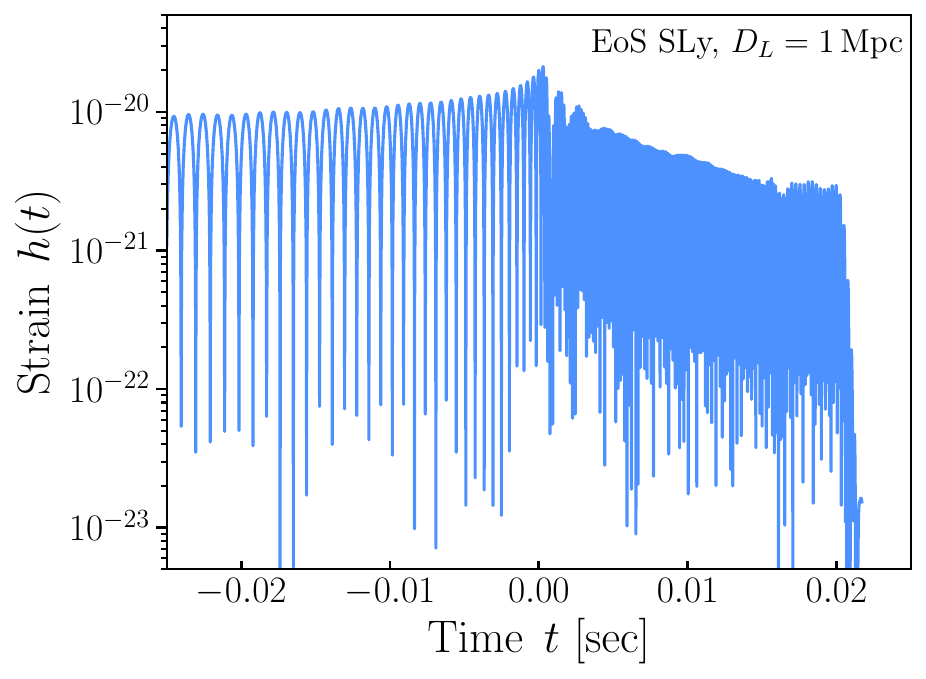}
	\hspace{0.3 cm}
	\includegraphics[width=0.48\textwidth]{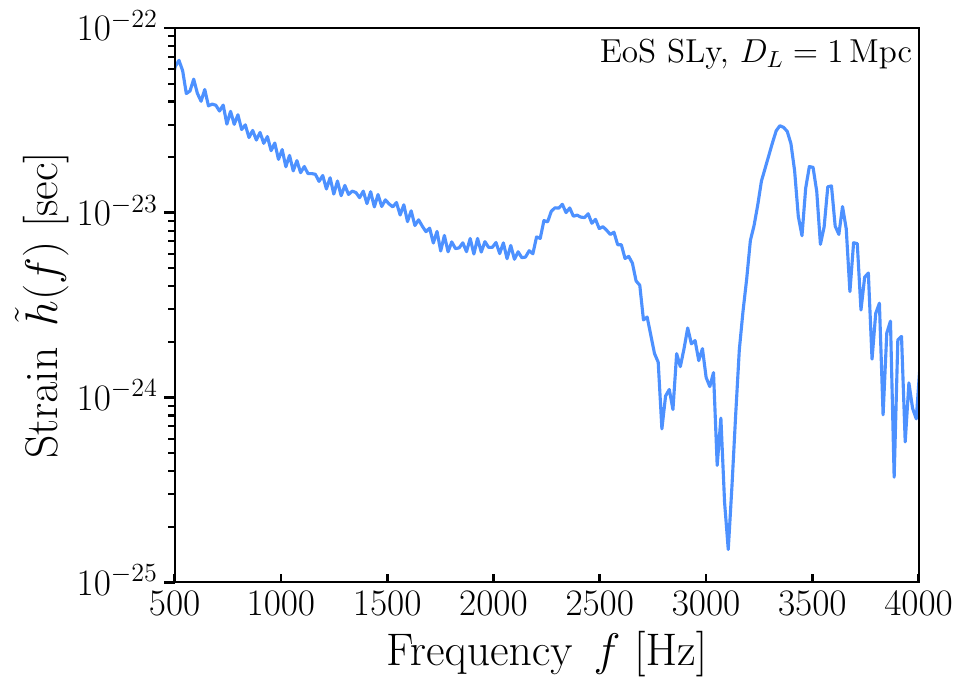}
	\caption{The time domain and frequency domain strain amplitudes are shown for the SLy EoS\,\mbox{\cite{Douchin:2001sv,Schneider:2017tfi}}, obtained from the BAM:0098 waveform dataset. Waveforms corresponding to seven additional EoSs have been extracted using the same procedure. For subsequent analysis, the frequency domain representations are used.}
	\label{fig_SLy}
\end{figure}

\subsection*{Strain vs Frequency Waveform from BAM:0098}

As an example, consider BAM:0098.h5 which is simulated with the EoS SLy\,\mbox{\cite{Douchin:2001sv,Schneider:2017tfi}}, with initial parameters as per our requirement. Fig.\,\ref{BAM:0098} shows its detailed data structure explaining the groups and attributes of the HDF5 file. The precise values of the initial input can be found in the metadata.txt file under the main file. We will now discuss the second group of the main data file, i.e., the $l=2,m=2$ strain part ($\rm rh_{22}$).

The dominant $l=m=2$ mode of the radiated GW, corresponds to $h_{22}(t)$ in the $rh_{22}$ part of the datafile. For this EoS we choose the \textnormal{BAM0098\_Rh\_l2\_m2\_r00800.txt} file, where the data is extracted at a radius of 800$M_{\odot}$ (in geometric units, with $G=c=1$, using mass as the dimension). The first three columns of this file provide the time ($u$), real part of $l=m=2$ strain (${\rm Re}\,rh_{22}$), imaginary part of $l=m=2$ strain (${\rm Im}\,rh_{22}$), respectively. The moment of the merger is defined as the time when the $h_{22}$ amplitude is largest. Waveforms are given in terms of retarded time\,\cite{Gonzalez:2022mgo},
\begin{align}
	u=t-r_{*}(r) = t-\left[r+r_s{\rm ln}\left(\frac{r}{r_s}-1\right)\right]\,,
\end{align}
where $r$ is the coordinate extraction radius in the simulations (= 800$M_{\odot}$, in this example), $r_{*}$ is the associated tortoise Schwarzschild coordinate, and $r_s= 2M$ is the Schwarzschild radius. From these inputs one can obtain the $h(t)$ vs. $t$ waveform. In Fig.\,\ref{fig_SLy} (left panel), we show the time domain waveform for face-on orientation ($\iota=0$) of the binary at a luminosity distance $D_{L}=1$\,Mpc. Here, $t=0$ sec denotes the time of merger. In the right panel of Fig.\,\ref{fig_SLy} the frequency domain $\tilde{h}(f)$ vs. $f$ waveform is shown, which is obtained using a Fourier transform on the time domain data.

\subsection{GW Detectors \emph{\&} their Exposure}
\label{sec:A2}
\change{We have used four detectors in this study whose sensitivity curve in the frequencies of our concern is shown in Fig.\ref{fig_Snvsf}.}

\paragraph{LIGO A+:} The upgraded version of Advanced LIGO (2.5-generation detector)\cite{LIGOScientific:2014pky} has the lowest sensitivity at both high and low frequencies, among the detectors we consider, limiting its ability to probe deep inspiral or postmerger dynamics compared to next-generation observatories. 

\paragraph{NEMO:}  NEMO (2.5-generation detector) is expected to operate at cryogenic temperatures and employs shorter, stiffer suspensions to reduce thermal vibrations, especially suspension thermal noise, a dominant source at high frequencies\,\cite{Ackley:2020atn}. Other key improvements include the reduction of internal thermal noise from the mirror substrates and coatings, as well as minimizing shot noise, which arises from quantum fluctuations in the laser light\,\cite{Ackley:2020atn}. These combined features make NEMO particularly well-suited for capturing the detailed structure of postmerger signals, enabling deeper insight into the NS EoS and other extreme matter phenomena.
	
\paragraph{Cosmic Explorer:}  CE ($3^{\rm rd}$-generation detector) is a proposed ground-based GW detector designed with 40 km arm lengths, an order of magnitude larger than LIGO's 4 km arms, offering vastly improved strain sensitivity across a broad frequency range, particularly in the low-frequency band ($5-1000$\,Hz), which is crucial for observing the full inspiral phase of CBC with high SNR\,\mbox{\cite{Reitze:2019iox, Evans:2021gyd}}.

\begin{figure}[t]
	\centering
	\includegraphics[width=0.5\textwidth]{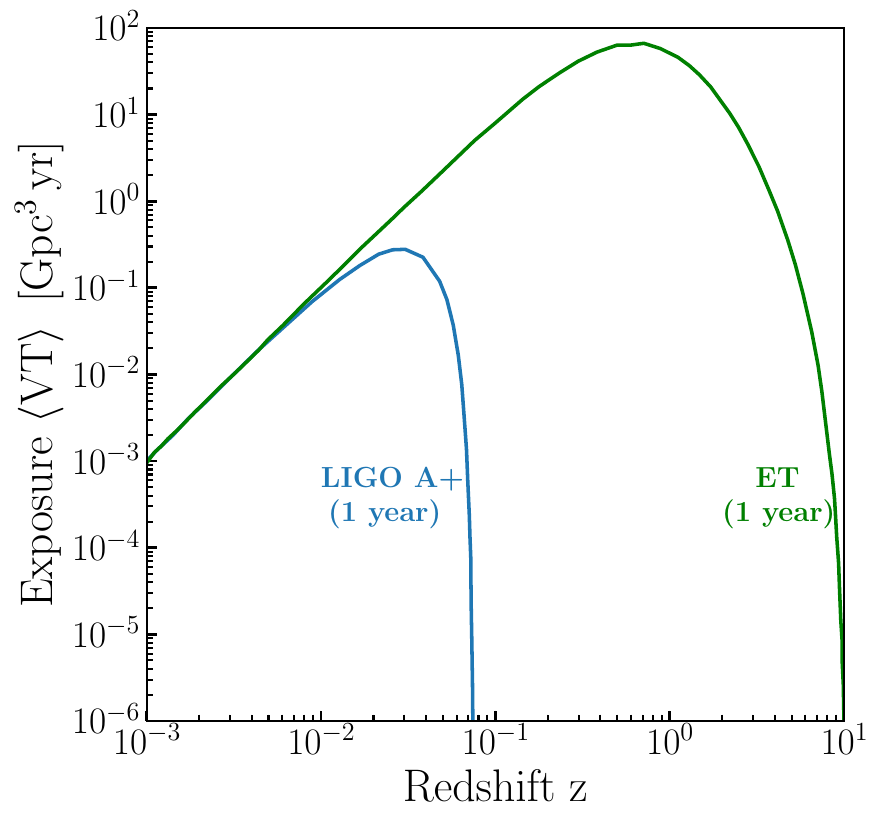}
	\caption{Exposure of the detectors. The detector specifications and the angular information are taken from\,\mbox{\cite{Taylor:2011fs, Taylor:2012db, Finn:1995ah}}. The value of $\mathcal{M}_c$ is chosen to be $1.17\,M_{\odot}$, i.e., the chirp mass of a symmetric binary with $m\,=\,1.35\,M_{\odot}$.}
	\label{fig_exposure}
\end{figure} 
	
\paragraph{Einstein Telescope:} ET ($3^{\rm rd}$-generation detector) is a proposed underground GW observatory featuring a unique triangular configuration with 10 km arms, designed to achieve unprecedented sensitivity across a broad frequency range, particularly below 10\,Hz, surpassing LIGO A+ and NEMO by mitigating low-frequency seismic, suspension and thermal noise through underground installation and cryogenic technologies\,\mbox{\cite{Punturo:2010zz, Abac:2025saz}}. ET and CE are particularly powerful for studying early inspiral physics, precision cosmology, and multi-band GW astronomy in synergy with space-based detectors. 

Here we present the exposure for LIGO A+ and the Einstein Telescope (Fig.\,\ref{fig_exposure}) that we have used for computing the event rates. The $ y $-axis shows the volume-time exposure, defined as the product of the sensitive volume and observation time,
\begin{align}
	\langle \rm VT \rangle&= T\times \int_0^{\infty} dz\, \frac{4\pi D_c^2(z)}{(1 + z) H(z)}\, C_\Theta\,
	\left[\frac{\rho_0}{8} 
	\left( \frac{D_L(z)}{r_0} \right)
	\left( \frac{1.2\, M_\odot}{(1 + z)\, \mathcal{M}_c} \right)^{5/6}\right]\,.
	\label{eq: exposure}
\end{align}
Here in Fig.\,\ref{fig_exposure}, we assume 1 year of data acquisition for each detector. The $ x $-axis corresponds to redshift. For reference, a redshift of $ z = 0.1 $ approximately corresponds to a luminosity distance ($D_L$) of $ \sim 450\,\mathrm{Mpc} $. The function $C_\Theta$ encodes the detection efficiency depending on mass, distance, and SNR of the source\,\mbox{\cite{Taylor:2011fs, Finn:1995ah}}. $D_L(z) = (1+z)D_c(z)$ is the luminosity distance at redshift $z$, and $\rho_0$ and $r_0$ denote the SNR threshold and characteristic distance reach of a particular detector, respectively. For our analysis we have considered  $\rho_0=8$, and two cases, $r_0=$ 80 Mpc  for LIGO A+ and 1591 Mpc for ET, respectively\,\cite {Finn:1995ah}.

\bibliographystyle{JHEP}
\bibliography{ref}	
\end{document}